\documentclass[12pt]{article}
\usepackage{amsfonts}
\usepackage{amssymb}
\usepackage{amsbsy}
\usepackage{mathrsfs}
\usepackage{amsmath}
\usepackage{graphicx}    
\usepackage{rotating}    
\usepackage{graphics}    
\usepackage{epsfig}
\usepackage{color}       

\usepackage[bottom]{footmisc}
\input epsf.sty

\newlength{\TZ}
\newcommand{\BEQ}{\begin{equation}}     
\newcommand{\BEA}{\begin{eqnarray}}
\newcommand{\BD}{\begin{displaymath}}
\newcommand{\EEQ}{\end{equation}}       
\newcommand{\EEA}{\end{eqnarray}}
\newcommand{\ED}{\end{displaymath}}
\newcommand{\bb}{\begin{eqnarray}}
\newcommand{\ee}{\end{eqnarray}}
\newcommand{\e}{{\rm e}}                
\newcommand{\vep}{\varepsilon}          
\newcommand{\D}{{\rm d}}                
\newcommand{\II}{{\rm i}}               
\renewcommand{\Re}{{\rm Re\ }}          

\newcommand{\sign}{{\rm sign\,}}        
\newcommand{\demi}{\frac{1}{2}}         
\newcommand{\wit}[1]{\widetilde{#1}}    
\newcommand{\wht}[1]{\widehat{#1}}      
\newcommand{\Tr}[1]{\operatorname{tr}\left( #1 \right)} 

\renewcommand{\vec}[1]{\boldsymbol{#1}} 

\newcommand{\vekz}[2]
     {\mbox{${\begin{array}{c} #1  \\ #2 \end{array}}$}}
\newcommand{\matz}[4]
     {\mbox{${\begin{array}{cc} #1 & #2 \\ #3 & #4 \end{array}}$}}

\newcommand{\appsection}[2]{\setcounter{equation}{0}\setcounter{subsection}{0}
\section*{Appendix #1. #2}
\renewcommand{\theequation}{#1.\arabic{equation}}
              \renewcommand{\thesection}{#1} }

\catcode`\@=11
\def\numberbysection{\@addtoreset{equation}{section}
        \def\theequation{\thesection.\arabic{equation}}}
\numberbysection

\definecolor{gruen}{rgb}{0,0.625,0}     
\definecolor{rot}{rgb}{0.75,0,0}        
\definecolor{blau}{rgb}{0,0,0.75}       

\begin{document}

\begin{titlepage}

\vskip 1.5 cm
\begin{center}
{\Large \bf Axiomatic construction of quantum Langevin equations}
\end{center}

\vskip 2.0 cm
\centerline{{\bf R\'ubia Ara\'ujo}$^{a,b}$, 
{\bf Sascha Wald}$^{c,}$\footnote{email: swald@sissa.it, ORCid:  0000-0003-1013-2130} and 
{\bf Malte Henkel}$^{a,d}$\footnote{courriel: malte.henkel@univ-lorraine.fr, Orcid: 0000-0002-5048-7852} 
}
\vskip 0.5 cm
\begin{center}
$^a$Laboratoire de Physique et Chimie Th\'eoriques (CNRS UMR 7019),\\  Universit\'e de Lorraine Nancy,
B.P. 70239, F -- 54506 Vand{\oe}uvre l\`es Nancy Cedex, France\\~\\
$^b$Departamento de Matem\'atica, Universidade Federal de Pernambuco,
Avenida Professor Moraes Rego, 1235 - Cidade Universit\'aria, CEP 50740-560, Recife - PE, Brazil\\~\\
$^c$SISSA - International School for Advanced Studies and INFN, via Bonomea 265, \\ I -- 34136 Trieste, Italia \\~\\    
$^d$Centro de F\'{i}sica Te\'{o}rica e Computacional, Universidade de Lisboa, \\P--1749-016 Lisboa, Portugal\\~\\
\end{center}

\begin{abstract}
A phenomenological construction of quantum Langevin equations, based on the physical criteria of 
(i) the canonical equal-time commutators, 
(ii) the Kubo formula, (iii) the virial theorem and (iv) the quantum fluctuation-dissipation theorem is presented. 
The case of a single harmonic oscillator coupled to a large external bath is analysed in detail. This allows to distinguish a
markovian semi-classical approach, due to Bedeaux and Mazur, from a non-markovian full quantum approach, 
due to to Ford, Kac and Mazur. The quantum-fluctuation-dissipation theorem is seen to be incompatible with a markovian dynamics. 
Possible applications to the quantum spherical model are discussed. 
\\~\\
\end{abstract}

\vfill

\end{titlepage}

\setcounter{footnote}{0}

\section{Introduction}

The description of quantum-mechanical many-body problems far from equilibrium 
presents conceptual difficulties which go beyond those present in classical
systems \cite{Breu02,Engl02,Cugl03,Gard04,Weis12,Taeu14}. 
In particular, while for classical systems either a master equation or a Langevin equation can readily be written down, 
this is far from obvious
for quantum systems. To  obtain at least formally a set of equations of motion, whose properties could then be studied, 
requires careful analysis of the quantum system. 
While this problem is quite well understood for isolated systems, 
for {\em open} quantum systems even the correct formulation of the problem appears to be not completely settled. 

In classical non-equilibrium statistical mechanics, 
writing down a Langevin equation is a standard, well-established approach  
for the description of the dissipative relaxation of any classical system towards its stationary (or equilibrium) state. 
However, its most straightforward extension to quantum systems does not work satisfactorily. 
We recall this difficulty using the example of a single harmonic oscillator,
see e.g. \cite{Carm99,Breu02}. If $s=s(t)$ denotes the magnetic `spin' variable\footnote{The oscillator coordinate 
$x=x(t)$ is labelled as a `spin' $s=s(t)$ in view of the planned applications to non-equilibrium quantum magnets.} 
and  $p=p(t)$ the canonically conjugated momentum, the `natural Langevin equation' would read
\BEQ \label{1.1}
\partial_t s = \frac{1}{m} p \;\; , \;\; \partial_t p = - m \omega^2 s - \lambda p +\eta
\EEQ 
where $m$ is the mass and $\omega$ the natural angular frequency of the oscillator. 
The coupling of the oscillator to an external heat bath is described by additional forces.  
These are (i) a dissipative force (assumed ohmic for simplicity) parametrised by the dissipation rate $\lambda$ and 
(ii) a random force, which is modelled by a centred gaussian noise $\eta$.  Consider the equal-time
commutator $c(t) =\langle[s(t),p(t)]\rangle$, where initially $c(0)=\II\hbar$. 
If $\langle [s(t),\eta(t)]\rangle=0$, eq.~(\ref{1.1}) implies the decay 
$c(t)=\II\hbar\, \e^{-t/\lambda}$ \cite{Carm99}. 
Such a behaviour is inconsistent with the one of a genuine {\em quantum} system,  even at temperature $T=0$. 
On the other hand,  eq.~(\ref{1.1}) does furnish an adequate treatment for classical dynamics. 
The problem of finding consistent {\em quantum} Langevin equations  ({\sc qle}s)
has been intensively discussed, see \cite{Gard04,Weis12} for detailed reviews and refs. therein. 

Bedeaux and Mazur \cite{Bede01,Bede02} suggested a different quantum Langevin equation.  
For the case of a single quantum harmonic oscillator ({\sc qho}), 
their proposal reads ($m,\omega,\lambda$ are positive constants)
\BEQ \label{1.2}
\partial_t s = \frac{1}{m} p +\eta_s \;\; , \;\; \partial_t p = - m \omega^2 s - \lambda p +\eta_p
\EEQ 
where the {\em two} distinct noises $\eta_p,\eta_s$ are assumed to be centred Gaussians, with the second moments
\begin{subequations} \label{1.3}
\begin{align} 
\left\langle \eta_p(t) \eta_p(t')\right\rangle 
&= \lambda m \hbar \omega \coth \left(\hbar \omega/2 k_{\rm B} T \right) \,\delta(t-t')         \label{1.3a} \\
\left\langle \eta_s(t) \eta_p(t')\right\rangle 
&= -\left\langle \eta_p(t) \eta_s(t')\right\rangle = \demi\, \II \hbar \lambda \,\delta(t-t')   \label{1.3b} \\
\left\langle \eta_s(t) \eta_s(t')\right\rangle 
&= 0                                                                                            \label{1.3c} 
\end{align}
\end{subequations}
and wherein the order of the operator noises $\eta_s,\eta_p$ is essential. 
Also, $T$ is the temperature of the bath and $\delta$ denotes the Dirac distribution. 
Clearly, eqs.~(\ref{1.2},\ref{1.3}) describe a system with the  Markov property. 
This approach was based on an analysis of the master equation of the
density matrix $\rho$, viz. $\hbar\partial_t \rho = -\II [H,\rho] + \hbar{\cal L}\delta S/\delta\rho$, 
where $S$ is the  thermodynamic entropy. The properties of the super-operator
${\cal L}$ are constrained by (i) the hermiticity of the thermodynamic force $\delta S/\delta\rho$, 
(ii) the conservation of probability, i.e.
$\Tr\rho=1$ for all times $t\geq 0$ and (iii) the Onsager relations. From this, 
they deduced the following equation of motion for time-dependent averages of an operator $A$ \cite{Bede01,Bede02}
\BEQ \label{1.4}
\frac{\D}{\D t} \left( \langle A\rangle(t) - \langle A\rangle_{\rm eq}\right) = \Tr{\frac{\D A}{\D t}\,\delta\rho}  
= \Tr{\left( \frac{\II}{\hbar}\left[ H, A\right] +\lambda A -\frac{\II \lambda}{\hbar} \left[ A p , x\right] \right)\delta\rho}
 \ ,
\EEQ
where $\delta\rho = \rho - \rho_{\rm eq}$ and $\rho_{\rm eq}$ 
denotes the equilibrium density matrix. They also worked out
all two-time Green's functions of spin and momentum in the stationary state from (\ref{1.4}) 
and verified that a direct solution of (\ref{1.2}) reproduces the noise correlators (\ref{1.3}). 

\begin{figure}[tb]
\begin{center}
\includegraphics[width=.5\hsize]{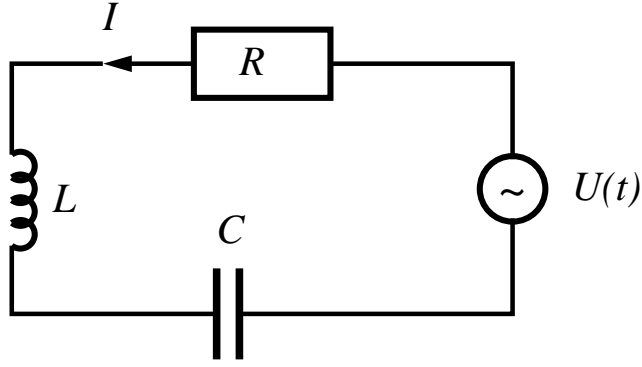}
\end{center}
\caption[fig0]{Schematic LRC circuit. 
\label{fig0} }
\end{figure}
{For illustration, we briefly mention two possible applications of the equations (\ref{1.2}).}
{{\bf (a)} First, consider a $LRC$ circuit, see fig.~\ref{fig0}.\footnote{For typical values of $L$ and $C$, 
the wavelength corresponding to the natural frequency $\sim \sqrt{LC\,}$ is much larger than typical sizes of nano-circuits; 
which are then in the so-called {\em lumped element limit} and can be described by a single degree of freedom \cite{Vool17}.} 
According to Kirchhoff's law, the total voltage is $U(t) = U_R+U_L+U_C$ and one may set $U_R=R I$ and $U_L = L \dot{I}$,
where $I=I(t)$ is the current. The coupling to an external bath is described by noise terms. The combined noises of the 
resistor $R$ and the coil $L$ are taken into account by setting $U(t)=L \eta_U$. For the capacitor $C$, the temporally rough changes in
the voltage, sometimes referred to as {\em `$kTC$-noise'} \cite{Johnson99}, are described by $\dot{U}_C = \frac{1}{C} I + \eta_I$. 
Herein, one will consider $\frac{1}{C}I(t)$ as the contribution to $\dot{U}_C$ which depends smoothly on the time $t$, 
whereas the noise $\eta_I(t)$ describes the rough contribution. 
Such a `reset noise' arises from the thermodynamic fluctuations of the amount of charge on the capacitor 
\cite{Becker78,Lundberg02,Sarpeshkar93,Gabe12,Vool17}.\footnote{Indeed, $C\eta_I=j$ is the current which arises from the 
random motions of the electric charges in the capacitor \cite[p. 295]{Becker78}. For quantum examples, see \cite{Gabe12}.}  
It is the dominant noise on sufficiently small capacitors and can be the limiting noise, 
e.g. in image sensors \cite{Lundberg02,Sarpeshkar93}. Combining the above equations leads to the system
\BEQ
\partial_t U_C = \frac{1}{C} I + \eta_I \;\; , \;\; 
\partial_t I = -\frac{R}{L} I - \frac{1}{L} U_C + \eta_U
\EEQ
which is identical to (\ref{1.2}), upon the correspondences $s \leftrightarrow U_C$, $p \leftrightarrow I$ and
$\eta_s \leftrightarrow \eta_I$, $\eta_p \leftrightarrow \eta_U$. While most of the literature discusses classical fluctuations, 
quantum fluctuations, as we shall discuss in this work, 
will become important for analysing noisy nano-electronic circuits \cite{Gabe12,Choi17,Crep17}. 
Their discussion may require to trace precisely the several potential sources of noise present. 
}

{{\bf (b)} Second, the oscillator (\ref{1.2}) also arises in model descriptions of the working of the inner ear of vertebrates 
\cite{Mart01,Dini12}. It is a known biological fact that the vertebrate ear not only admits but also emits sound, the so-called
{\em spontaneous otoacoustic emissions}. Such an active process in the inner ear is well-described by models near to an oscillatory
instability which occurs on the onset of a Hopf bifurcation \cite{Mart01}. Specialised mechanoreceptors, i.e. bundles of hair cells, 
provide the work required of such active processes. Indeed, eq.~(\ref{1.2}) is a special case of the most simple of such models, 
where $s$ describes the hair-bundle position and $p$ the force exerted by the active process. The two noises $\eta_s, \eta_p$, 
respectively, describe the effects of fluctuations on the bundle position and the active process, respectively \cite{Mart01,Dini12}. 
}

How can one decide whether a given system of equations of motion, or a master equation, can describe a genuine quantum system~? 
Using the proposal (\ref{1.2},\ref{1.3}) of Bedeaux and Mazur as a phenomenological device, 
we formulate the following {\em minimal criteria} for a reasonable description of quantum dynamics: 
\begin{enumerate}
\item[{\bf (A)}] the {(averaged)} canonical equal-time commutators 
$\bigl\langle [s_n(t), p_m(t)]\bigr\rangle=\II\hbar\, \delta_{n,m}$ 
should be kept for all times $t> 0$, if initially obeyed. 
\item[{\bf (B)}] the Kubo formul{\ae} of linear response theory should be reproduced 
(we shall recall the precise definition in section~2). 
\end{enumerate}
As we shall see, these requirements will be sufficient to fix the commutators between the two noises $\eta_s,\eta_p$. 
In addition, one must require that 
\begin{enumerate}
\item[{\bf (C)}] the virial theorem should be reproduced (it is valid in both classical and quantum mechanics and is used to
analyse strongly interacting many-body systems at `thermal' equilibrium. 
{Important examples arise in astrophysics and cosmology for
self-gravitating systems, where commonly `virialisation' and `equilibration' are treated as synonyms \cite{Daux02,Harw06,Schn08,Mo10}}). 
\item[{\bf (D)}] the quantum fluctuation-dissipation theorem ({\sc qfdt}) 
will be an essential characteristic of quantum equilibrium states. 
{It has been recognised since a long time that the
{\sc qfdt} is the key requirement for a stationary state being a quantum equilibrium state, 
in order to be consistent with the second fundamental
theorem of quantum thermodynamics \cite{Hang05,Ford17}.} 

Indeed, it has been understood recently that the {\sc qfdt} is
a consequence of the invariance of the Keldysh path integral under 
a combined time-reversal and Kubo-Martin-Schwinger transformation \cite{Sieb15,Aron18}. 
Moving away from the quantum equilibrium state will break this symmetry and the {\sc qfdt} will cease to be valid. Therefore,
the confirmation of the {\sc qfdt} will serve to distinguish quantum equilibrium states from any other type of stationary state.  
\end{enumerate} 
These two conditions will fix the anti-commutators of the two noises $\eta_s,\eta_p$. 
Of course, a discussion must include the question whether
all these requirements can be satisfied simultaneously. 

Here, we wish to explore which kinds of dynamics are compatible with these criteria.  
{In any case, a microscopic derivation of a `quantum Langevin equation' which describes thermalisation should obey a 
series of physically reasonable consistency requirements, which we take to be the four criteria {\bf (A,B,C,D)}.} 
We shall proceed in several steps. 
In section~2, we shall re-consider the case of a single harmonic oscillator, 
described by (\ref{1.2}) following the suggestion of \cite{Bede01,Bede02} of including two
noises $\eta_s,\eta_p$, but we shall also include the effect of initial conditions. 
Proceeding from the exact solution of (\ref{1.2}), we shall show that the phenomenological criteria {\bf (A,B,C)} formulated
above  provide increasing constraints on the two noises $\eta_s,\eta_p$ 
such that the combination of all three criteria finally reproduces the proposal
(\ref{1.3}) of Bedeaux and Mazur. On the other hand, the requirement {\bf (D)} 
of the {\sc qfdt} only holds true in the $\hbar\to 0$ limit. 
In section~3, a proposal for a quantum Langevin equation of the damped harmonic oscillator, 
with two noises $\eta_s,\eta_p$, will be presented such that all four
criteria {\bf (A,B,C,D)} will be obeyed. In addition, the noise correlators are independent of the model parameters $m,\omega$ 
and only contain the damping constant $\lambda$. 
Also, the noise correlators are non-markovian. A possible explicit form is  
\begin{subequations} \label{1.5}
\begin{align} 
\left\langle \eta_s(t) \eta_s(t')\right\rangle &= \left\langle \eta_p(t) \eta_p(t')\right\rangle =0  \label{1.5a} \\
\left\langle \eta_s(t) \eta_p(t')\right\rangle &=  -\left\langle \eta_p(t) \eta_s(t')\right\rangle = 
\demi\, \II \hbar \lambda\, \delta(t-t') +\demi \lambda k_{\rm B} T \coth\left(\frac{\pi k_{\rm B} T}{\hbar} (t-t')\right)    
\label{1.5b} 
\end{align}
\end{subequations}
where the order of the operators is essential. We shall see that the non-markovianity is a direct 
consequence of the {\sc qfdt}. 
Eqs.~(\ref{1.2},\ref{1.5}) state our proposal for the quantum Langevin equation of the damped quantum harmonic oscillator, 
in the special case of ohmic damping. Since the noise correlators do not depend on the parameters $m,\omega$ 
of the model we used as a scaffold to derive them, they should be generic and applicable in wider contexts. 
We shall also show that the {\sc qle}s (\ref{1.2},\ref{1.5}) are equivalent to the second-order Ford-Kac-Mazur ({\sc fkm}) 
quantum Langevin equation \cite{Ford65,Ford87,Ford88,Gard04,Weis12} of a particle in an external potential.  

We must add that eqs.~(\ref{1.5}) are not the only solution to the criteria {\bf (A,B,C,D)}. Rather, we shall see that the
most general solution can be expressed by two anti-symmetric functions $\psi(t),\chi(t)$ and two symmetric functions
$\alpha(t),\beta(t)$ which also depend on the parameters $m,\omega$. 

In section~4, we shall illustrate a few simple consequences of the noise correlators (\ref{1.5}). 
The Langevin equations of the quantum harmonic oscillator can be re-interpreted as the {\sc qle} 
of the modes of the {\em quantum spherical model} in $d$ spatial dimensions. The non-trivial 
physics (i.e. distinct from mean-field theory) 
of this model follows from the coupling of these modes by the so-called {\em spherical constraint}, which
is included into the {\sc qle} via a time-dependent Lagrange multiplier.  
The equilibrium critical behaviour of the quantum spherical model is well-known 
\cite{Ober72,Henk84a,Niew95,Vojt96,Oliv06,Sach11,Bien12,Bien13,Wald15}: 
while for finite temperatures $T>0$, the critical behaviour is
the same as for the classical spherical model \cite{Berl52,Lewi52}, at temperature $T=0$ there is a quantum critical
point in the same universality class as the $(d+1)$-dimensional classical spherical model \cite{Kogu79}.  
The time-dependent non-equilibrium classical system has also been very thoroughly analysed, 
see e.g. \cite{Ronc78,Coni94,Godr00b,Fusc02,Pico02,Cugl03,Hase06,Ebbi08,Dura17}. 
The non-equilibrium behaviour of {\em isolated} quantum or classical spherical models was studied recently \cite{Scio13,Mara15,Cugl17}. 
The long-time dynamics of a disordered quantum spherical model was analysed in the mean-field limit \cite{Rokn04}. 
The quantum dynamical behaviour of the {\em open} quantum spherical model quenched to temperature $T=0$ 
as described by a Lindblad equation was recently studied \cite{Wald16,Wald18,Wald18a,Timp18} and it was shown that for
quantum quenches deep into the ordered state, 
the long-time behaviour of several averages becomes independent of the coupling to the bath.  
{We shall outline how the explicit solution of the quantum spherical dynamics can be considerably simplified 
in the long-time limit via a scaling argument. This can be done for both the Bedeaux-Mazur noise correlators (\ref{1.3}) 
as well as for the quantum noise correlators (\ref{1.5}).} 
The qualitative difference of quantum and classical  correlators will also be illustrated for the {\sc qho}. 
{We have therefore explicitly formulated the {\sc qle} of the quantum spherical model. 
However, finding the late-time solutions explicitly is still technically demanding and will be left to future work.} 
We conclude in section~5. Several appendices give further background and calculational details.

\section{Semi-classical dissipative harmonic oscillator}

The quantum Langevin equations (\ref{1.2}) are characterised by the two noises $\eta_s(t)$ and $\eta_p(t)$. 
In addition, we shall have to characterise the fluctuations of the initial state. 
We shall apply the criteria formulated in section~1 to constrain 
their possible form. From now on, we set $k_{\rm B}=1$. 

\subsection{Formal solution}

The first step of the solution of eq.~(\ref{1.2}) is to re-write the homogeneous part in matrix form
\BEQ
\partial_t \left( \vekz{s}{p} \right) 
= \left(\matz{0}{1/m}{-m \omega^2}{-\lambda}\right) \left( \vekz{s}{p} \right) = -\Lambda \left( \vekz{s}{p} \right)
\EEQ
The eigenvalues of the matrix $\Lambda$ are
\BEQ \label{2.2}
\Lambda_{\pm} = \frac{\lambda}{2} \pm \sqrt {\frac{\lambda^2}{4} -\omega^2\,}
\EEQ
and they obey the following properties: if $0<|\omega|< \lambda/2$, then $\Lambda_+>\Lambda_{-}>0$. 
If $|\omega|> \lambda/2>0$, then $\Lambda_{-}^*=\Lambda_+$ and $\Re \Lambda_{\pm}=\lambda/2>0$. 

The homogeneous solution implies the following ansatz 
\BEQ \label{2.3}
s(t) = s_+(t) \e^{-\Lambda_+ t} + s_{-}(t) e^{-\Lambda_- t} \;\; , \;\; p(t) = p_+(t) \e^{-\Lambda_+ t} + p_{-}(t) e^{-\Lambda_- t}
\EEQ
in order to solve eq. (\ref{1.2}) with the method of variation of constants.
Two of the four amplitudes ($s_{\pm}, p_{\pm}$) in the ansatz can be still chosen freely. Our choice is
\BEQ \label{2.4}
p_{\pm}(t) = - m \Lambda_{\pm} s_{\pm}(t)
\EEQ
Now, we reinsert the ansatz (\ref{2.3}) into the eqs.~(\ref{1.2}), and use the conditions (\ref{2.2},\ref{2.4}). 
This leads to the simplified system (the dots denote the derivative with respect to time) 
\BEQ
\dot{s}_+(t) \e^{-\Lambda_+ t} + \dot{s}_{-}(t) \e^{-\Lambda_{-} t} = \eta_s(t) \;\; , \;\;
-\Lambda_+ \dot{s}_+(t) \e^{-\Lambda_+ t} - \Lambda_{-} \dot{s}_{-}(t) \e^{-\Lambda_{-} t} = \frac{1}{m} \eta_p(t)
\EEQ
which can be separated as follows (throughout, we assume $|\omega|\ne \lambda/2$) 
\BEQ
\dot{s}_+(t) = - \frac{\exp (\Lambda_+ t)}{\Lambda_+ - \Lambda_{-}} \left( \frac{\eta_p}{m} + \Lambda_{-} \eta_s \right) \;\; , \;\;
\dot{s}_-(t) =  \frac{\exp (\Lambda_- t)}{\Lambda_+ - \Lambda_{-}} \left( \frac{\eta_p}{m} + \Lambda_{+} \eta_s \right)
\EEQ
Their immediate integration then gives the formal exact solution of (\ref{1.2})
%
%
%
%
 \begin{subequations} \label{2.7}
 \begin{align} 
 s(t) &= s_+(0) \e^{-\Lambda_+ t} + s_{-}(0) \e^{-\Lambda_{-} t} \nonumber \\
      & \hspace{-0.7truecm}-\frac{1}{\Lambda_+ - \Lambda_-} 
      \int_0^t \!\D\tau\: \e^{-\Lambda_{+} (t-\tau)} \left[ \frac{\eta_p(\tau)}{m} + \Lambda_{-}\eta_s(\tau) \right]
      +\frac{1}{\Lambda_+ - \Lambda_-} 
      \int_0^t \!\D\tau\: \e^{-\Lambda_{-} (t-\tau)} \left[ \frac{\eta_p(\tau)}{m} + \Lambda_{+}\eta_s(\tau) \right]
      \label{2.7a} \\
 p(t) &= -m \Lambda_+ s_+(0) \e^{-\Lambda_+ t} -m \Lambda_{-}  s_{-}(0) \e^{-\Lambda_{-} t} \nonumber \\
      & \hspace{-0.7truecm}+\frac{m \Lambda_{+}}{\Lambda_+ - \Lambda_-} 
      \int_0^t \!\D\tau\: \e^{-\Lambda_{+} (t-\tau)} \left[ \frac{\eta_p(\tau)}{m} + \Lambda_{-}\eta_s(\tau) \right]
      -\frac{m \Lambda_{-}}{\Lambda_+ - \Lambda_-} 
      \int_0^t \!\D\tau\: \e^{-\Lambda_{-} (t-\tau)} \left[ \frac{\eta_p(\tau)}{m} + \Lambda_{+}\eta_s(\tau) \right]
      \label{2.7b} 
 \end{align}
 \end{subequations}
Eqs.~(\ref{2.7}) will be the basis of all subsequent calculations. 

\subsection{Equal-time commutator}

While the amplitudes $s_{\pm}(0)$ characterise the fluctuations of the initial state, 
the fluctuations of the bath are described by $\eta_s(t)$ and
$\eta_p(t)$. We assume that these two types of fluctuations are independent of each other and therefore that on average
\BEQ \label{gl:aux_comm}
\left\langle \left[ s_{\pm}(0), \eta_a(t) \right] \right\rangle = 0 \;\; , \;\;
\left\langle \left[ \eta_{a}(t), \eta_{a}(t') \right]\right\rangle = 0
\EEQ
where $a=s,p$ labels the two noises. The second relation asserts that noises of the same kind 
(either for the spin or for the momentum) should commute. 
It remains to fix the commutator $\left\langle \left[ s_+(0), s_{-}(0) \right]\right\rangle$ and we also consider the mixed commutator
\BEQ \label{2.9}
\left\langle \left[ \eta_{s}(t) , \eta_p(t') \right]\right\rangle = \kappa \, \delta(t-t')
\EEQ
where the constant $\kappa$ is to be found. Here, we shall use the $\delta$-distribution in order to simplify the calculations. 
The physical argument behind is that we observe the system's dynamics on time-scales large with respect 
to any correlation times in the bath. 
If that assumption should not be justified, a different bath correlation kernel must be used 
(in that case $\kappa\delta(t-t')$ would become a time-dependent function $\kappa(t-t')$). 
We shall not explore that possibility in this section, however. 
 
Using the exact solution (\ref{2.7}), the spin-momentum commutator is readily worked out
\begin{align}
\left\langle\big[ s(t), p(t') \big]\right\rangle &= 
- \frac{\kappa}{(\Lambda_{+}-\Lambda_{-})(\Lambda_{+}+\Lambda_{-})} 
\bigg(\Lambda_{+} \e^{-\Lambda_{-}|t-t'|} - \Lambda_{-} \e^{-\Lambda_{+}|t-t'|} \bigg)
\label{2.10} \\
& \hspace{-0.9truecm}+m \bigg( \Lambda_+ \e^{-\Lambda_{-} t -\Lambda_{+} t'} - \Lambda_- \e^{-\Lambda_{+} t -\Lambda_{-} t'} \bigg)
\bigg( \left\langle\big[ s_{+}(0), s_{-}(0) \big]\right\rangle 
- \frac{\kappa /m}{(\Lambda_{+}-\Lambda_{-})(\Lambda_{+}+\Lambda_{-})}\bigg)
\nonumber 
\end{align}
We identify two contributions: for large enough times, only the stationary part will contribute 
(the first  line in (\ref{2.10})), 
but there is also a non-stationary part, given by the second line of (\ref{2.10}). The stationarity
of the spin-momentum commutator can be achieved by requiring the following initial condition
\BEQ \label{gl:sp-komm}
\left\langle\left[ s_{+}(0), s_{-}(0) \right]\right\rangle =  \frac{\kappa /m}{(\Lambda_{+}-\Lambda_{-})(\Lambda_{+}+\Lambda_{-})}
\EEQ
The sought canonical commutator is obtained from the remaining stationary commutator 
$\left\langle\left[ s(t), p(t') \right]\right\rangle$ when we also set $t=t'$. We read off 
\BEQ
\left\langle\left[ s(t), p(t) \right]\right\rangle =  \frac{\kappa}{\lambda} \stackrel{!}{=} \II \hbar
\EEQ
so that we can identify
\BEQ \label{gl:can_comm}
\kappa = \II\lambda \hbar \;\; , \;\; 
\left\langle \left[ \eta_s(t), \eta_p(t') \right]\right\rangle = \II\lambda \hbar\, \delta(t-t')
\EEQ
The second of these reproduces eq.~(\ref{1.3b}) as proposed in \cite{Bede01}. 

\begin{figure}[tb]
\begin{center}
\includegraphics[width=.5\hsize]{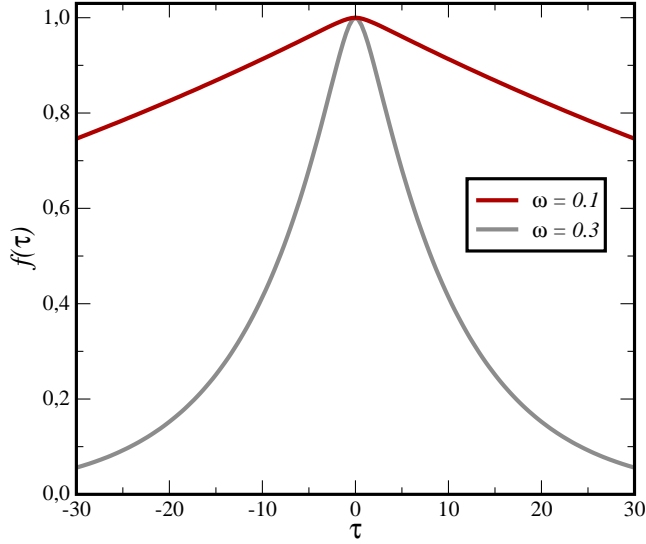}
\end{center}
\caption[fig1]{Time-dependence of the normalised commutator $f(\tau)=\langle [s(\tau),p(0)]\rangle/(\II\hbar)$, 
over against the time difference $\tau$, for $\lambda=1$ and two values of $\omega$. 
\label{fig1} }
\end{figure}
On the other hand, for different times $t\ne t'$, the canonical spin-momentum commutator is not kept, but only depends on
the time difference $\tau=t-t'$. The functional form of the two-time commutator is illustrated in figure~\ref{fig1}.
For increasing values of $\omega$ (but still small enough that $\Lambda_{\pm}$ are real), the peak at $\tau=0$ becomes more sharp 
but there is no simple and model-independent argument to predict this form.  

\subsection{Linear responses} 

As a preparation for the discussion of the Kubo formula and the fluctuation-dissipation theorem ({\sc fdt}), 
we compute responses of the average spin and momentum to an external magnetic perturbation. 
The hamiltonian reads $H=\frac{p^2}{2m}+\frac{m\omega^2}{2} s^2 - h s $. The quantum equations of motion of the {\em closed} system
are $\dot{p}=\frac{\II}{\hbar}[H,p]$ and $\dot{s}=\frac{\II}{\hbar}[H,s]$, 
where the equal-time canonical commutator $[s,p]=\II\hbar$ is to be used.
Adding to these the damping and noise terms of the open system, the full equations of motion become now 
\BEQ
\dot{s} = \frac{1}{m} p +\eta_s \;\; , \;\; \dot{p} = - m \omega^2 s + h - \lambda p  +\eta_p
\EEQ
The linear responses of the spin $s$ and of the momentum $p$ are defined as
\BEQ
R^{(s)}(t,t') := \left. \frac{\delta \langle s(t)\rangle}{\delta h(t')}\right|_{h=0} \;\; , \;\; 
Q^{(p)}(t,t') := \left. \frac{\delta \langle p(t)\rangle}{\delta h(t')}\right|_{h=0}
\EEQ
They obey the equations
\BEQ \label{gl:reponses_h}
\partial_t R^{(s)}(t,t') = \frac{1}{m} Q^{(p)}(t,t') \;\; , \;\;
\partial_t Q^{(p)}(t,t') = -m \omega^2 R^{(s)}(t,t') -\lambda Q^{(p)}(t,t') +  \delta(t-t')
\EEQ
Writing $\mathscr{R}(t-t',t')=R^{(s)}(t,t')$ and denoting $\tau=t-t'$ yields
\BEQ
\partial_{\tau}^2 \mathscr{R}(\tau,t') +\lambda \partial_{\tau} \mathscr{R}(\tau,t') 
+ \omega^2 \mathscr{R}(\tau,t') =\delta(\tau)/m 
\EEQ
Hence, the response $R^{(s)}=R^{(s)}(\tau)=\mathscr{R}(\tau,t')$ is stationary and does not explicitly depend
on $t'$ alone. Introducing the Fourier transform 
$\wht{R}^{(s)}(\nu) = (2\pi)^{-1/2} \int_{\mathbb{R}} \!\D\tau\: \e^{-\II\nu \tau} R^{(s)}(\tau)$ leads to the stationary solution
\BEQ \label{gl:Rs-nu}
\wht{R}^{(s)}(\nu) = -\frac{1}{m\sqrt{2\pi\,}} \frac{1}{\nu^2-\II\lambda\nu - \omega^2} 
\EEQ
$\wht{R}^{(s)}(\nu)$ has two simple poles in the upper complex $\nu$-plane. The spin response function $R^{(s)}$ reads
\BEQ \label{gl:Rs}
R^{(s)}(t-t') = \Theta(t-t')\frac{1/m}{\Lambda_{+}-\Lambda_{-}}\left(\e^{-\Lambda_{-} (t-t')}-\e^{-\Lambda_{+} (t-t')}\right) 
\EEQ
where the Heaviside function $\Theta$ expresses the causality condition $t\geq t'$. 
The magnetic momentum response $Q^{(p)}(t,t')$ can be read off from (\ref{gl:reponses_h}). 

Analogously, one can consider a perturbation by the momentum operator, 
where now $H=H=\frac{p^2}{2m}+\frac{m\omega^2}{2} s^2 + k p$. The equations of motion now read
\BEQ
\dot{s} = \frac{1}{m} p + k + \eta_s \;\; , \;\; \dot{p} = - m \omega^2 s - \lambda p  +\eta_p 
\EEQ
and the linear response functions are now defined as 
\BEQ
R^{(p)}(t,t') := \left. \frac{\delta \langle p(t)\rangle}{\delta k(t')}\right|_{k=0} \;\; , \;\; 
Q^{(s)}(t,t') := \left. \frac{\delta \langle s(t)\rangle}{\delta k(t')}\right|_{k=0} 
\EEQ
An analogous calculation as before leads to the (stationary) momentum response function
\BEQ \label{gl:Rp}
R^{(p)}(t-t')  = \frac{m \omega^2\, \Theta(t-t')}{\Lambda_{+}-\Lambda_{-}} 
\left( \e^{-\Lambda_{-} (t-t')} -  \e^{-\Lambda_{+} (t-t')}\right) = m^2 \omega^2 R^{(s)}(t-t') 
\EEQ

\subsection{Commutators and the Kubo formul{\ae}} 

In terms of the density matrix $\rho(0)$, time-dependent averages of an operator $A(t)$ 
are defined as $\langle A(t) \rangle = Z^{-1}\Tr{A(t)\rho(0)}$ where
$Z:=\Tr{\rho(0)}$. Following \cite{Cugl03}, two-time correlators are defined as follows
\begin{subequations} \label{gl:correl}
\begin{align} 
C_{+}(t,t') &:= \demi \left\langle A(t) B(t') + B(t') A(t)\right\rangle 
= \demi \left\langle \left\{ A(t), B(t')\right\} \right\rangle \label{gl:corr_s} \\
C_{-}(t,t') &:= \demi \left\langle A(t) B(t') - B(t') A(t)\right\rangle 
= \demi \left\langle \left[ A(t), B(t')\right]   \right\rangle \label{gl:corr_a}
\end{align}
\end{subequations}
In what follows, we shall admit in these definitions that $A(t)=B(t)$, unless stated explicitly otherwise. 
Our choices will be $A=s$ or $A=p$, respectively. 
Then the {\em Kubo formul{\ae}} \cite{Cugl03,Weis12}
\BEQ \label{gl:Kubo} 
R^{(s,p)}(t-t') = \frac{2\II}{\hbar} \Theta(t-t')\, C_{-}^{(s,p)}(t,t') 
\EEQ
relate the anti-symmetrised correlator $C_{-}(t,t')$ to the linear response function of an operator $A(t)$ 
with respect to its conjugate field. 
Note that the commutator $C_{-}(t,t')$ is antisymmetric in $t,t'$ and we shall also see explicitly its stationarity. 
The Kubo formul{\ae} are consequences of merely the linear response. However, their validity is not related 
to the condition that the system which is perturbed had to be at equilibrium,
nor that the hamiltonian should be conserved on average. 

We proceed to check (\ref{gl:Kubo}) for the spin and momentum responses in the model at hand. 
First, consider the anti-symmetrised spin-spin correlator, which is evaluated using (\ref{2.7a}) 
\begin{align}\nonumber
C_{-}^{(s)}(t,t') &= \demi  \left\langle \left[ s_{+}(0), s_{-}(0)\right]\right\rangle 
\left( \e^{-\Lambda_+ t - \Lambda_- t'} - \e^{-\Lambda_- t - \Lambda_+ t'}\right) 
\nonumber \\
&  +\frac{1/2}{(\Lambda_+ -\Lambda_-)^2} \int_{0}^{t}\!\D\tau\int_0^{t'}\!\D\tau'\left( \e^{-\Lambda_+(t-\tau + t'-\tau')} 
\bigg<\bigg[ \frac{\eta_{p}(\tau)}{m} + \Lambda_{-} \eta_{s}(\tau), \frac{\eta_{p}(\tau')}{m} 
+ \Lambda_{-} \eta_{s}(\tau') \bigg] \bigg> \right. 
\nonumber \\
&  \hspace{3.75cm}- \e^{-\Lambda_+(t-\tau)} \e^{-\Lambda_{-}(t'-\tau')} 
\left\langle \left[ \frac{\eta_{p}(\tau)}{m} + \Lambda_{-} \eta_{s}(\tau), \frac{\eta_{p}(\tau')}{m} 
+ \Lambda_{+} \eta_{s}(\tau') \right] \right\rangle 
\nonumber \\
&  \hspace{3.75cm}- \e^{-\Lambda_-(t-\tau)} \e^{-\Lambda_{+}( t'-\tau')} 
\left\langle \left[ \frac{\eta_{p}(\tau)}{m} + \Lambda_{+} \eta_{s}(\tau), \frac{\eta_{p}(\tau')}{m} 
+ \Lambda_{-} \eta_{s}(\tau') \right] \right\rangle 
\nonumber \\
&  \left. \hspace{3.75cm} +\ \e^{-\Lambda_-(t-\tau + t'-\tau')} 
\left\langle \left[ \frac{\eta_{p}(\tau)}{m} + \Lambda_{+} \eta_{s}(\tau), 
\frac{\eta_{p}(\tau')}{m} + \Lambda_{+} \eta_{s}(\tau') \right] \right\rangle  \right)
\nonumber 
\end{align}

These commutators are evaluated by using (\ref{gl:aux_comm},\ref{gl:can_comm}). 
The first and fourth commutator vanish and by using the
initial condition (\ref{gl:sp-komm}), the other two simplify as follows
\begin{align}\nonumber 
C_{-}^{(s)}(t,t') &= \frac{\lambda \hbar}{\II m} \frac{1/2}{\Lambda_+ - \Lambda_-} 
\bigg\{ -\frac{ \e^{-\Lambda_+ t - \Lambda_- t'} - \e^{-\Lambda_- t - \Lambda_+ t'}}{\Lambda_+ +\Lambda_-} -
\int_0^{\min(t,t')} \!\D\tau\: \e^{-\Lambda_+ t-\Lambda_- t'} \e^{(\Lambda_+ +\Lambda_-)\tau} \nonumber \\
&   \hspace{7.95cm}+
\int_0^{\min(t,t')} \!\D\tau\: \e^{-\Lambda_- t-\Lambda_+ t'} \e^{(\Lambda_+ +\Lambda_-)\tau} \bigg\}
\nonumber \\
&= \frac{\hbar}{2\II m} \frac{1}{\Lambda_+ - \Lambda_-}\: \mbox{\rm sign\,} (t-t') \:
\left( \e^{-\Lambda_- |t-t'|} - \e^{-\Lambda_+ |t-t'|} \right)
\label{gl:C_as}
\end{align}
We point out the stationarity of this anti-symmetrised correlator, 
which is a consequence of the assumptions made before on the noise and initial commutators. 
Technically, the stationarity comes about since after integration, the contributions from the
lower integration limit cancel the terms coming from the initial condition. 

Second, we consider the anti-symmetrised momentum-momentum correlator, 
which is computed using (\ref{2.7b}). In the same manner as before, we find
\BEA
C_{-}^{(p)}(t,t') = \demi \left\langle \left[ p(t), p(t') \right]\right\rangle
= \frac{m\hbar}{2\II} \frac{\Lambda_+ \Lambda_-}{\Lambda_+ - \Lambda_-}\: \mbox{\rm sign\,} (t-t')
\left( \e^{-\Lambda_- |t-t'|} - \e^{-\Lambda_+ |t-t'|} \right)
\label{gl:C_ap}
\EEA

In order to compare these forms with the response functions, recall from (\ref{2.2}) 
that $\Lambda_+ + \Lambda_-=\lambda$ and $\Lambda_+ \Lambda_- = \omega^2$. 
Then, letting $\tau=t-t'>0$ and comparing eqs.~(\ref{gl:C_as},\ref{gl:Rs}) and (\ref{gl:C_ap},\ref{gl:Rp}), respectively, we see that
\BEQ
R^{(s)}(t-t') = \frac{2\II}{\hbar} \Theta(t-t')\, C_{-}^{(s)}(t-t') \;\; , \;\; 
R^{(p)}(t-t') = \frac{2\II}{\hbar} \Theta(t-t')\, C_{-}^{(p)}(t-t')
\EEQ
which confirms the Kubo formul{\ae} (\ref{gl:Kubo}) for the spin and momentum responses. 
Although these response functions and commutators are stationary, this
does {\em not} necessarily mean that the underlying stochastic process must be stationary itself, 
since much freedom still remains for the choice of the initial conditions. 

Summarising, we have shown: {\em if for the linearly damped harmonic oscillator with equations of motion (\ref{1.2}) 
and the two noises $\eta_s(t)$ and $\eta_p(t)$, the
noise correlator (\ref{gl:can_comm}) (and also (\ref{gl:aux_comm})) and the initial condition (\ref{gl:sp-komm}), namely
$\left\langle \left[ s_+(0), s_-(0)\right]\right\rangle = (\II\hbar/2m) \left[ \lambda^2/4-\omega^2\right]^{-1/2}$, 
are obeyed, then 
\begin{enumerate}
\item for all times $t\geq 0$, the averaged equal-time commutator 
$\left\langle \left[ s(t), p(t)\right]\right\rangle = \II\hbar$ is canonical
\item the Kubo formul{\ae} (\ref{gl:Kubo}) hold true for all times $t>t'>0$. 
\end{enumerate}
} 

\noindent 
This gives a sufficient condition to satisfy the criteria {\bf (A)} and {\bf (B)}, formulated in section~1. 
In addition, we have seen that
$R^{(p)}(\tau)/R^{(s)}(\tau)=C_{-}^{(p)}(\tau)/C_{-}^{(s)}(\tau)=m^2\omega^2$, for the damped harmonic oscillator. 

\subsection{Anti-commutators and the virial theorem} 

In order to characterise the stationary state further, the anti-commutators of the noises are required. 
As suggested by \cite{Bede01,Bede02}, the ansatz
\BEA
\left\langle \eta_p(t) \eta_p(t')\right\rangle = \demi \left\langle \left\{ \eta_p(t), \eta_p(t') \right\}\right\rangle 
= \alpha \delta(t-t') 
\nonumber \\
\left\langle \eta_s(t) \eta_s(t')\right\rangle = \demi \left\langle \left\{ \eta_s(t), \eta_s(t') \right\}\right\rangle 
= \beta \delta(t-t') 
\label{2.28}
\EEA
will be tried, along with the natural assumption $\left\langle \left\{ \eta_s(t), \eta_p(t') \right\}\right\rangle = 0$. 
Herein, the constants $\alpha,\beta$  must be found. From (\ref{2.7}), we obtain the spin-spin 
correlator\footnote{Herein and below, we denote the
averages over the bath, as well as the one over the initial conditions, by the same symbol $\langle \cdot \rangle$, 
although these two averages are unrelated.} 
\begin{align}\nonumber
C_{+}^{(s)}(t,t') = \left[  \left\langle s_+(0)^2\right\rangle - \frac{\frac{\alpha}{m^2}
+\beta\Lambda_-^2}{\Lambda_+ (\Lambda_+ - \Lambda_-)^2}\right] \frac{\e^{-\Lambda_+(t+t')} }{2}
+ \left[ \left\langle s_-(0)^2\right\rangle - \frac{\frac{\alpha}{m^2} +\beta\Lambda_+^2}{\Lambda_- (\Lambda_+ - \Lambda_-)^2}\right] 
\frac{\e^{-\Lambda_-(t+t')} }{2}
\nonumber \\
 + \bigg[ \frac{\left\langle \left\{ s_+(0), s_-(0)\right\}\right\rangle }{2}
+ \frac{\frac{\alpha}{m^2} +\beta\Lambda_+\Lambda_-}{(\Lambda_+ +\Lambda_-) (\Lambda_+ - \Lambda_-)^2}   \bigg]
\bigg[\e^{-\Lambda_+ t - \Lambda_- t'}+ \e^{-\Lambda_+ t' - \Lambda_- t} \bigg]
\nonumber \\
  +\bigg[\frac{\frac{\alpha}{m^2}+\textcolor{black}{\Lambda^2_-}\beta}{2\Lambda_+}
  -\frac{\frac{\alpha}{m^2}+\textcolor{black}{\Lambda_+\Lambda_-}\beta}{2(\Lambda_++\Lambda_-)}\bigg]
 \frac{\e^{-\Lambda_+|t-t'|} }{(\Lambda_+-\Lambda_-)^2} 
 +\bigg[\frac{\frac{\alpha}{m^2}+\textcolor{black}{\Lambda_+^2}\beta}{2\Lambda_-} 
 -\frac{\frac{\alpha}{m^2}+\textcolor{black}{\Lambda_+\Lambda_-}\beta}{2(\Lambda_++\Lambda_-)}\bigg]
    \frac{\e^{-\Lambda_-|t-t'|} }{(\Lambda_+-\Lambda_-)^2}
\label{gl:corrS}
\end{align}
%
We observe stationary terms (in the last line) and rapidly decaying non-stationary contributions (whenever $t,t'\gg |t-t'|$), 
which solely depend on the initial conditions. 
Similarly, the momentum-momentum correlator reads
\begin{align}\nonumber
C_{+}^{(p)}(t,t') =&\:  m^2 \Lambda_+^2 \left[ \left\langle s_+(0)^2\right\rangle 
  - \frac{\frac{\alpha}{m^2} +\beta\Lambda_-^2}{\Lambda_+ (\Lambda_+ - \Lambda_-)^2}\right]  \frac{\e^{-\Lambda_+(t+t')}}{2}
\nonumber \\
& + m^2 \Lambda_-^2 \left[ \left\langle s_-(0)^2\right\rangle 
  - \frac{\frac{\alpha}{m^2} +\beta\Lambda_+^2}{\Lambda_- (\Lambda_+ - \Lambda_-)^2}
\right]  \frac{\e^{-\Lambda_-(t+t')} }{2}
\nonumber \\
& + m^2 \Lambda_+\Lambda_- \left[ \frac{\left\langle \left\{ s_+(0), s_-(0)\right\}\right\rangle}{2} 
+ \frac{\frac{\alpha}{m^2} + \beta \Lambda_+\Lambda_-}{(\Lambda_+ +\Lambda_-) (\Lambda_+ - \Lambda_-)^2}  \right]
\left[\e^{-\Lambda_+ t - \Lambda_- t'}+ \e^{-\Lambda_+ t' - \Lambda_- t} \right]
\nonumber \\
&  +\frac{m^2 \Lambda_+^2\left(\frac{\alpha}{m^2}+\Lambda_-^2\beta\right)}{2\Lambda_+(\Lambda_+-\Lambda_-)^2}\e^{-\Lambda_+|t-t'|} 
    +\frac{m^2 \Lambda_-^2\left(\frac{\alpha}{m^2}+\Lambda_+^2\beta\right)}{2\Lambda_-(\Lambda_+-\Lambda_-)^2}\e^{-\Lambda_-|t-t'|} 
\nonumber \\
&  -\frac{m^2 \Lambda_+\Lambda_- \left(\frac{\alpha}{m^2}+\Lambda_+\Lambda_-\beta\right)}{(\Lambda_++\Lambda_-)(\Lambda_+-\Lambda_-)^2} 
    \left( \e^{-\Lambda_+|t-t'|} + \e^{-\Lambda_-|t-t'|} \right)
\label{gl:corrP}
\end{align}
and has an analogous structure. 
The virial theorem relates these two correlators at equilibrium, namely (see appendix~A)
\BEQ
\left\langle p^2 \right\rangle_{\rm eq} = m^2 \omega^2 \left\langle s^2 \right\rangle_{\rm eq} \ .
\EEQ
Therefore, we must require that
\BEQ
C_{+}^{(p)}(t,t)_{\rm stat} \stackrel{!}{=} m^2 \Lambda_+ \Lambda_- C_{+}^{(s)}(t,t)_{\rm stat} \ .
\EEQ
Therein, the last line from eq.~(\ref{gl:corrP}) cancels against contributions in (\ref{gl:corrS})
and the remaining stationary terms lead to the condition
\begin{align*}
\Lambda_+ \left( \frac{\alpha}{m^2} + \beta \Lambda_-^2\right) +  \Lambda_- \left( \frac{\alpha}{m^2} 
+ \beta \Lambda_+^2\right)
\stackrel{!}{=} 
\Lambda_- \left( \frac{\alpha}{m^2} + \beta \Lambda_-^2\right) + \Lambda_+ \left( \frac{\alpha}{m^2} 
+ \beta \Lambda_+^2\right) \ .
\end{align*}
For $\beta=0$, this is indeed an identity. For $\beta\ne 0$, the terms containing $\alpha$ cancel and using the explicit form of 
$\Lambda_{\pm}$, we finally arrive at $\lambda^3 \stackrel{!}{=} 4\lambda \omega^2$. However, 
since $\lambda,\omega$ are independent model parameters, 
a generic solution with $\beta\ne 0$ is impossible. The result $\beta=0$ reproduces (\ref{1.3c}). 

Having shown that $\beta=0$, it remains to find $\alpha$. The spin-spin correlator simplifies to 
\begin{align}\nonumber 
C_{+}^{(s)}(t,t')=& \left[\left\langle s_+(0)^2\right\rangle - 
\frac{{\alpha}/{m^2}}{\Lambda_+ (\Lambda_+ - \Lambda_-)^2}\right] 
\frac{\e^{-\Lambda_+(t+t')}}{2}
+ \left[ \left\langle s_-(0)^2\right\rangle - 
\frac{{\alpha}/{m^2}}{\Lambda_- (\Lambda_+ - \Lambda_-)^2}\right]
\frac{\e^{-\Lambda_-(t+t')}}{2}
\nonumber \\
&+ \left[ \demi \left\langle \left\{ s_+(0), s_-(0)\right\}\right\rangle 
+ \frac{{\alpha}/{m^2} }{(\Lambda_+ +\Lambda_-) (\Lambda_+ - \Lambda_-)^2}   \right]
\left[\e^{-\Lambda_+ t - \Lambda_- t'}+ \e^{-\Lambda_+ t' - \Lambda_- t} \right]
\nonumber \\
& + \frac{\alpha}{2\lambda m^2 \omega^2} \frac{\Lambda_+ \e^{-\Lambda_- |t-t'|} - \Lambda_- \e^{-\Lambda_+ |t-t'|}}{\Lambda_+-\Lambda_-}
\label{gl:corrS2}
\end{align}
The equilibrium contribution is the stationary part (third line of (\ref{gl:corrS2})) at $t=t'$. 
For the quantum harmonic oscillator at thermal equilibrium 
it is explicitly known that
\BEQ \label{2.34}
C_+^{(s)}(t,t)_{\rm stat} = \frac{\alpha}{2\lambda m^2 \omega^2} \stackrel{!}{=} 
\frac{\hbar}{2m\omega} \coth \frac{\hbar \omega}{2 T} = \left\langle s^2\right\rangle_{\rm eq} 
\EEQ
hence
\BEQ \label{2.35}
\alpha = \hbar \lambda m \omega \coth \frac{\hbar \omega}{2 T}
\EEQ
as expected from (\ref{1.3a}). Clearly, in the limit $\hbar\to 0$ one goes back to the classical result. 
Therefore, {\em the criteria {\bf (A,B,C)} as specified in section~1, together with the ans\"atze (\ref{2.9},\ref{2.28}), 
reproduce the proposal (\ref{1.3}) of Bedeaux and Mazur for the noises $\eta_s,\eta_p$}. 

\subsection{On the fluctuation-dissipation relationship} 

Given the explicit results of the previous sub-sections, we can now inquire about the relationship between the stationary parts of the 
correlators $C_+^{(s,p)}$ and the responses $R^{(s,p)}$. Empirically, we find in the stationary state, using eq. (\ref{2.35}) 
\BEQ \label{gl:fdt-BM}
\partial_{\tau} C_+^{(s)}(\tau)_{\rm stat} = - \hbar \omega \coth \left(\frac{\hbar \omega}{2 T}\right) R^{(s)}(\tau) \;\; , \;\;
\partial_{\tau} C_+^{(p)}(\tau)_{\rm stat} = - \hbar \omega \coth \left(\frac{\hbar \omega}{2 T}\right) R^{(p)}(\tau) 
\EEQ
First, eq.~(\ref{gl:fdt-BM}) illustrates that the relationship between responses and correlators does not depend on the observable. 
Second, these relations do {\em not} correspond to the habitual 
{\em quantum} fluctuation-dissipation theorem ({\sc fdt}), 
although in the limit $\hbar\to 0$, the standard {\em classical} {\sc fdt} is recovered. 
This means that from the minimal criteria as formulated above in  section~1, only the criteria {\bf (A,B,C)} 
are obeyed and indeed lead back to the proposal eqs.~(\ref{1.2},\ref{1.3}), 
while {\bf (D)} can only be satisfied in the $\hbar\to 0$ limit.\footnote{{In view of recent results \cite{Kubo18} 
which assert that the {\sc qfdt} does not hold if quasi-classical measurments are made, 
semi-classical models such as the present one might acquire some independent interest.}} 
Hence {\em the proposal (\ref{1.3}) for the noise(s) leads at best to a semi-classical description of the relaxation.} 

This example explicitly illustrates that the requirement of the canonical commutator relations, 
in spite of reproducing the Kubo formul{\ae}, by itself
is not enough to guarantee that the stationary state of the dynamics is truly a quantum equilibrium state. 
Also, the noise correlators (\ref{1.3}) depend explicitly
on the model parameters $m, \omega$ such that the recipe proposed in \cite{Bede01,Bede02} 
is explicitly model-dependent and can only be applied to systems
of harmonic oscillators. More generally, the ansatz studied in this section assumed 
$\delta$-correlated noise from the outset. While this certainly
simplifies calculations, the physical origin of that assumption remains unclear and we have seen that from such an ansatz the 
{\sc qfdt} {\em cannot} be derived. 

\section{Quantum dissipative harmonic oscillator}

We now seek to generalise the approach of the previous section such that all four criteria {\bf (A,B,C,D}) are obeyed. 
Physically, the noises $\eta_s,\eta_p$ describe the properties of the heat bath and its coupling to the system. 
If the bath is very large with respect to the system and bath correlations decay on a much faster time-scale than 
the typical system's time-scale, its
properties will be independent of the system's behaviour and the bath can be considered stationary. 
Hence, the correlators of $\eta_s,\eta_p$ can be
found from the stationary behaviour of the system coupled to the heat bath. 
Then the analysis can be simplified by going over to frequency space, via the Fourier transformation
\BEQ \label{gl:Fourier}
(\mathscr{F} s(t) )(\nu) = \wht{s}(\nu) := \frac{1}{\sqrt{2\pi\,}\,} \int_{\mathbb{R}} \!\D t\: e^{-\II \nu t} s(t) 
\EEQ
The equations of motion (\ref{1.2}) become (for ohmic damping with $\lambda>0$)
\BEQ
\II\nu \wht{s}(\nu) = \frac{1}{m} \wht{p}(\nu) + \wht{\eta}_s(\nu) \;\; , \;\;
\II\nu \wht{p}(\nu) =  -m \omega^2 \wht{s}(\nu) - \lambda \wht{p}(\nu) +\wht{\eta}_p(\nu)
\EEQ
Hence, the formal solution is
\begin{subequations} \label{gl:solution}
\begin{align} 
\wht{s}(\nu) &= \frac{\wht{\eta}_p(\nu)/m + (\II\nu+\lambda)\wht{\eta}_s(\nu)}{\omega^2+\II\lambda\nu-\nu^2} \label{gl:solution_s} \\
\wht{p}(\nu) &= \frac{\II\nu \wht{\eta}_p(\nu) - m \omega^2 \wht{\eta}_s(\nu)}{\omega^2+\II\lambda\nu-\nu^2} \label{gl:solution_p}
\end{align}
\end{subequations}
This will be the basis of all following calculations. 

\subsection{Equal-time commutator} 

In order to characterise the equal-time commutator $\left\langle\bigl[s(t),p(t)\bigr]\right\rangle$, 
the commutators of the two noises must be analysed. We assume
\BEQ \label{3.4} 
\left\langle \bigl[ \eta_s(t), \eta_s(t') \bigr]\right\rangle =\left\langle \bigl[ \eta_p(t), \eta_p(t') \bigr]\right\rangle = 0 
\;\; , \;\;
\left\langle \bigl[ \eta_s(t), \eta_p(t') \bigr]\right\rangle = \II \hbar \kappa(t-t')
\EEQ 
The form of the third commutator, is motivated by the system-plus-bath being stationary. 
In appendix~C, we show that requiring the conditions (\ref{3.4}), where in addition the yet undetermined function 
$\kappa(t)=\kappa(-t)$ must be symmetric,  
is the most simple way to guarantee the validity of the Kubo formul{\ae}. 
In frequency-space, we have (using (\ref{gl:Fourier})) 
\BEQ \label{3.5} 
\left\langle \bigl[ \wht{\eta}_s(\nu) , \wht{\eta}_p(\nu') \bigr]\right\rangle 
= \delta(\nu+\nu') \II\hbar \sqrt{2\pi\,}\, \wht{\kappa}(\nu)
\EEQ
Similarly, for the spin-momentum commutator, we expect the stationary form
\BEQ \label{3.6}
\left\langle \bigl[ \wht{s}(\nu), \wht{p}(\nu') \bigr]\right\rangle = \delta(\nu+\nu') \II\hbar \sqrt{2\pi\,}\, \wht{K}(\nu)
\;\; , \;\; 
\left\langle \bigl[ {s}(t), {p}(t') \bigr]\right\rangle = \II\hbar  {K}(t-t')
\EEQ
where $K(t)$ must be found such that $K(0)\stackrel{!}{=}1$ reproduces the required equal-time canonical commutator, 
at least on average, according to criterion {\bf (A)}. 

Inserting the formal solution (\ref{gl:solution}) into (\ref{3.6}) and using (\ref{3.4}) gives
\BEQ \label{3.7}
\wht{K}(\nu) = \frac{\nu^2 -\II\lambda \nu +\omega^2}{(\nu^2 -\II\lambda\nu-\omega^2)(\nu^2 +\II\lambda\nu-\omega^2)}\: 
\wht{\kappa}(\nu)
\EEQ
Then $K(t) = (f\star \kappa)(t)$ is a convolution, where the Fourier transform $\wht{f}(\nu)$ can be read off from (\ref{3.7}). 
If we assume that $\wht{\kappa}(\nu)=\kappa_0$ is a constant, we obtain $K(0)=\sqrt{2\pi\,}\,\kappa_0 \lambda^{-1}$ 
such that the normalisation requirement $K(0)=1$ implies
\BEQ
\sqrt{2\pi\,}\,\kappa_0 = \lambda
\EEQ
We then have the non-vanishing noise commutators
\BEQ \label{gl:bruit-commutateur}
\left\langle \bigl[ \eta_s(t), \eta_p(t') \bigr]\right\rangle = \II\hbar\lambda \delta(t-t') \;\; , \;\;
\left\langle \bigl[ \wht{\eta}_s(\nu), \wht{\eta}_p(\nu') \bigr]\right\rangle = \delta(\nu+\nu') \II\hbar\lambda
\EEQ
Under the simplifying assumptions made, this commutator is the same as proposed by Bedeaux and Mazur \cite{Bede01,Bede02}.  
Indeed, a markovian form of the noise commutator appears to be natural for the chosen ohmic dissipation. 

\subsection{Linear responses} 

Adding perturbations to the closed-system Hamiltonian according to $H = p^2/2m + \demi m \omega^2 s^2 - h s + k p$, 
the response functions are the same as calculated in section~2. From (\ref{gl:Rs-nu},\ref{gl:Rp}), we have
\BEQ \label{gl:Rsp-fourier} 
\wht{R}^{(s)}(\nu) = -\frac{1}{m\sqrt{2\pi\,}} \frac{1}{\nu^2-\II\lambda\nu - \omega^2} = \frac{1}{m^2 \omega^2} \wht{R}^{(p)}(\nu)
\EEQ
for the linear responses of spin and momentum  with respect to their conjugate fields. 

\subsection{Kubo formul{\ae}} 

\begin{figure}[tb]
\begin{center}
\includegraphics[width=.5\hsize]{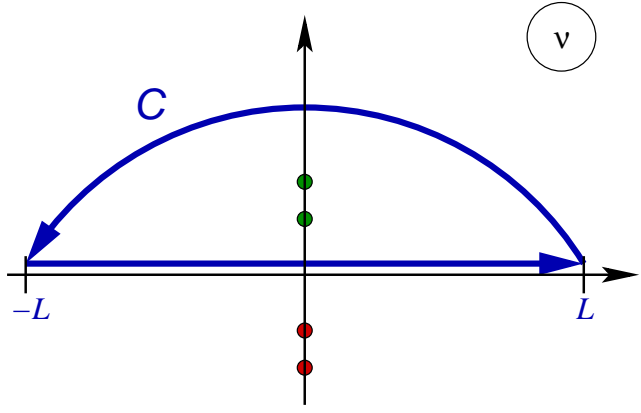}
\end{center}
\caption[fig2]{Integration contour $C$ for the verification of the Kubo formula {\protect (\ref{3.13})}. 
The poles $\nu_{+,\pm}$ ($\nu_{-,\pm}$) in the upper (lower) complex $\nu$-plane are indicated, for $|\omega|<\lambda/2$, 
by green filled (red open) circles. 
At the end, the limit $L\to\infty$ is taken. 
\label{fig2} }
\end{figure}

Using eq. (\ref{gl:solution}), we have
\begin{subequations} \label{3.11}
\begin{align}
C_{-}^{(s)}(\nu,\nu') &= \demi \left\langle\left[ \wht{s}(\nu), \wht{s}(\nu') \right]\right\rangle = \delta(\nu+\nu') 
\wht{C}_{-}^{(s)}(\nu) 
\\
C_{-}^{(p)}(\nu,\nu') &= \demi \left\langle\left[ \wht{p}(\nu), \wht{p}(\nu') \right]\right\rangle = \delta(\nu+\nu') 
\wht{C}_{-}^{(p)}(\nu) 
\end{align}
\end{subequations}
where
\BEQ \label{3.12}
\wht{C}_{-}^{(s)}(\nu) = - \frac{\hbar\lambda}{m} \frac{\nu}{(\nu^2-\II\lambda\nu-\omega^2)(\nu^2+\II\lambda\nu-\omega^2)} 
= \frac{1}{m^2 \omega^2}\, \wht{C}_{-}^{(p)}(\nu) 
\EEQ
Since the Heaviside function is non-local in frequency-space, we cannot compare this directly with the responses (\ref{gl:Rsp-fourier}). 
Rather, in order to check the Kubo formul{\ae}\ae~(\ref{gl:Kubo}), let $\tau=t-t'>0$ 
and examine in frequency-space the validity of the Kubo formula in the form (\ref{gl:A:Kubo-Fourier}), that is
\BEQ \label{3.13}
\frac{2\II}{\hbar}\, C_{-}^{(s)}(\tau) \Theta(\tau) = R^{(s)}(\tau) \stackrel{?}{=}  
-\frac{2}{\hbar\II}\, \frac{1}{2\pi} \int_{\mathbb{R}} \!\D\nu\: \e^{\II\nu\tau}\, \wht{C}_{-}^{(s)}(\nu)
\EEQ
where $\wht{C}_{-}^{(s)}(\nu)$ is taken from (\ref{3.12}). In the complex $\nu$-plane, this function has simple poles at 
$\nu_{\pm,\pm}=\pm \frac{\II\lambda}{2}\pm \II \sqrt{ \lambda^2/4-\omega^2\,}$, 
see figure~\ref{fig2}. Of these, $\nu_{+,\pm}$ are in the upper half-plane and
$\nu_{-,\pm}$ are in the lower half-plane. In addition, at the poles $\nu=\nu_{+,\pm}$, one has
\BEQ \label{3.14} 
\left.\frac{\nu}{\nu^2+\II\lambda\nu-\omega^2}\right|_{\nu=\nu_{+,\pm}} = \frac{1}{2\II\lambda}
\EEQ
When calculating the integral in (\ref{3.13}) via the residue theorem, see fig.~\ref{fig2}, 
the contour $C$ will be closed in the upper half-plane since $\tau>0$ and only the
residua at $\nu_{+,\pm}$ will contribute. Therefore, in order to check (\ref{3.13}), 
consider the contour integral (see fig.~\ref{fig2}), with $\tau>0$ 
\BEA 
R^{(s)}(\tau) &\stackrel{?}{=}& \frac{2\II}{\hbar}\, \frac{1}{2\pi} \oint_{C} \!\D\nu\: \e^{\II\nu\tau} \frac{\hbar\lambda}{m}
\frac{\nu}{(\nu^2+\II\lambda\nu-\omega^2)}\frac{-1}{(\nu^2-\II\lambda\nu-\omega^2)} 
\nonumber \\
&=&  \frac{2\II}{\hbar}\, \frac{1}{2\pi} \frac{\hbar\lambda}{m} \frac{1}{2\II\lambda} 
\oint_{C} \!\D\nu\: \e^{\II\nu\tau} \frac{-1}{(\nu^2-\II\lambda\nu-\omega^2)} 
\nonumber \\
&=& \frac{1}{\sqrt{2\pi\,}\,} \int_{\mathbb{R}} \!\D\nu\: \e^{\II\nu\tau} \frac{1}{m\sqrt{2\pi\,}\,} 
\frac{-1}{(\nu^2-\II\lambda\nu-\omega^2)} 
\nonumber \\
&=& \frac{1}{\sqrt{2\pi\,}\,} \int_{\mathbb{R}} \!\D\nu\: \e^{\II\nu\tau} \wht{R}^{(s)}(\nu)
\EEA
as required. Herein, in the first line the commutator $\wht{C}_{-}^{(s)}(\nu)$ was inserted from (\ref{3.12}); 
in the second line the residua at $\nu=\nu_{+,\pm}$ 
were computed according to  (\ref{3.14}) and finally the response function (\ref{gl:Rsp-fourier}) was recognised. 
Of course, the contribution of the upper semi-arc is negligible. 

This shows the validity of the Kubo formula (\ref{gl:A:Kubo},\ref{gl:A:Kubo-Fourier}) for the spin $s$. 
Its validity for the momentum $p$ is now obvious from (\ref{gl:Rsp-fourier}) and (\ref{3.12}). 

\textcolor{blau}{\underline{\textcolor{black}{Summarising:}}} 
{\em if for an ohmic dissipation, the only non-vanishing noise correlators are given by (\ref{gl:bruit-commutateur}), 
then (i) for all times $t\geq 0$ the equal-time canonical commutator $\left\langle \left[ s(t), p(t)\right]\right\rangle=\II\hbar$ 
and (ii) for all time differences $\tau=t-t'>0$ the Kubo formul{\ae} for both the spin $s$ and the
momentum $p$ hold true.} This implements the criteria {\bf (A,B)}. 

\subsection{Anti-commutators and the virial theorem} 

It remains to analyse the admissible forms of the noise anti-correlators. Since the bath is stationary, we can start from
\BEQ
\left\langle \left\{ \eta_s(t), \eta_s(t') \right\}\right\rangle = 2 \beta(t-t') \ , \
\left\langle \left\{ \eta_p(t), \eta_p(t') \right\}\right\rangle = 2 \alpha(t-t') \ , \
\left\langle \left\{ \eta_s(t), \eta_p(t') \right\}\right\rangle = 2 \gamma(t-t') 
\EEQ
where the functions $\alpha(t)=\alpha(-t)$, $\beta(t)=\beta(-t)$ and $\gamma(t)$ must be found. In frequency-space this reads
\BEA
\left\langle \left\{ \wht{\eta}_s(\nu), \wht{\eta}_s(\nu') \right\}\right\rangle &=& \delta(\nu+\nu') 
\sqrt{2\pi\,}\: 2 \wht{\beta}(\nu)  \nonumber \\
\left\langle \left\{ \wht{\eta}_p(\nu), \wht{\eta}_p(\nu') \right\}\right\rangle &=& \delta(\nu+\nu') 
\sqrt{2\pi\,}\: 2 \wht{\alpha}(\nu)  \\
\left\langle \left\{ \wht{\eta}_s(\nu), \wht{\eta}_p(\nu') \right\}\right\rangle &=& \delta(\nu+\nu') 
\sqrt{2\pi\,}\: 2 \wht{\gamma}(\nu) \nonumber  
\EEA
While $\wht{\alpha}(\nu)$ and $\wht{\beta}(\nu)$ are symmetric, there is no obvious symmetry for $\wht{\gamma}(\nu)$. 
With the definition
\begin{subequations}
\begin{align}
C_{+}^{(s)}(\nu,\nu') &= \demi\left\langle\left\{ \wht{s}(\nu), \wht{s}(\nu')\right\}\right\rangle 
= \delta(\nu+\nu') \wht{C}_{+}^{(s)}(\nu)  \\
C_{+}^{(p)}(\nu,\nu') &= \demi\left\langle\left\{ \wht{p}(\nu), \wht{p}(\nu')\right\}\right\rangle 
= \delta(\nu+\nu') \wht{C}_{+}^{(p)}(\nu) 
\end{align}
\end{subequations}
we have from (\ref{gl:solution}) 
\begin{subequations} \label{gl:correlateurs}
\begin{align} 
\wht{C}_{+}^{(s)}(\nu) &:= 
\frac{\frac{\wht{\alpha}(\nu)}{m^2} +(\lambda^2+\nu^2)\wht{\beta}(\nu)
+\frac{\lambda}{m}\left( \wht{\gamma}(\nu)+\wht{\gamma}(-\nu)\right)
+\frac{\II\nu}{m}\left(\wht{\gamma}(\nu)-\wht{\gamma}(-\nu)\right)}{(\nu^2-\II\lambda\nu-\omega^2)(\nu^2+\II\lambda\nu-\omega^2)
} 
\label{gl:correlateur_s} \\[.25cm]
\wht{C}_{+}^{(p)}(\nu) &:= 
\frac{\nu^2\wht{\alpha}(\nu) + m^2 \omega^4 \wht{\beta}(\nu)
+\II m \omega^2 \nu \left(\wht{\gamma}(\nu)-\wht{\gamma}(-\nu)\right)}{(\nu^2-\II\lambda\nu-\omega^2)(\nu^2+\II\lambda\nu-\omega^2)
} 
\label{gl:correlateur_p}
\end{align}
\end{subequations}
The virial theorem states that $\wht{C}_{+}^{(p)}(\nu)\stackrel{!}{=}m^2 \omega^2 \wht{C}_{+}^{(s)}(\nu)$, see appendix~A. 
Combination with (\ref{gl:correlateurs}) leads to the condition
\begin{align}
(\nu^2 - \omega^2)\wht{\alpha}(\nu) + (m^2 \omega^4 -   m^2 \omega^2 (\lambda^2 + \nu^2))\wht{\beta}(\nu) 
\stackrel{!}{=} \lambda m \omega^2 \left( \wht{\gamma}(\nu) + \wht{\gamma}(-\nu)\right)
\end{align}
The simplest solution is 
\BEQ \label{gl:taux-abc}
\wht{\alpha}(\nu) = \wht{\beta}(\nu) = 0 \;\; , \;\; \wht{\gamma}(\nu) + \wht{\gamma}(-\nu) = 0 
\EEQ
It is the only solution which does not depend on the model's parameters $m,\omega$. 
Any other solution must contain at least one of them. Since we are looking
for a characterisation of the noises $\eta_s,\eta_p$ which should be independent of the specific physical model under study, 
we shall retain the solution (\ref{gl:taux-abc}) as our implementation of the condition {\bf (C)}. 

Inserting (\ref{gl:taux-abc}), the correlators take the form 
\BEQ \label{3.22}
\wht{C}_{+}^{(s)}(\nu) = 
\frac{2\II}{m}\frac{\nu \wht{\gamma}(\nu)}{(\nu^2-\II\lambda\nu-\omega^2)(\nu^2+\II\lambda\nu-\omega^2)} 
= \frac{1}{m^2 \omega^2}\, \wht{C}_{+}^{(p)}(\nu) 
\EEQ

\subsection{Quantum fluctuation-dissipation theorem}

The form of the remaining antisymmetric function $\wht{\gamma}(\nu)=-\wht{\gamma}(-\nu)$ is found from the {\sc qfdt}. 
As shown in appendix~A, in frequency-space it can be written in the form
\BEQ
\frac{\wht{C}_{+}^{(s)}(\nu)}{\wht{C}_{-}^{(s)}(\nu)} = \frac{\wht{C}_{+}^{(p)}(\nu)}{\wht{C}_{-}^{(p)}(\nu)} 
= - \frac{1}{\sqrt{2\pi\,}\,} \coth\frac{\hbar \nu}{2T}
\EEQ
Therefore, comparison of (\ref{3.12}) with (\ref{3.22}) leads to the condition
\BEQ \label{gl:gamma}
2\II \sqrt{2\pi\,}\, \wht{\gamma}(\nu) = \hbar \lambda \coth\frac{\hbar \nu}{2 T}
\EEQ
which also implements the condition {\bf (D)}. 
This means that the only non-vanishing commutators and anti-commutators of the noises are
\BEQ \label{3.25} 
\left\langle\left\{ \wht{\eta}_s(\nu), \wht{\eta}_p(\nu') \right\}\right\rangle 
= \frac{\hbar\lambda}{\II} \coth\left( \frac{\hbar\nu}{2T}\right)\, \delta(\nu+\nu') \;\; , \;\;
\left\langle\left[ \wht{\eta}_s(\nu), \wht{\eta}_p(\nu') \right]\right\rangle 
=\II\hbar\lambda \, \delta(\nu+\nu')
\EEQ
They do depend on the assumption of ohmic dissipation and on the properties of the bath (here the temperature $T$). 
But since they do not contain the specific parameters $m,\omega$ of the harmonic oscillator model we used to derive them, 
they should be applicable in general. 

It is possible to combine the equations of motion (\ref{1.2}) into a single second-order Langevin equation, 
which reads \cite{Ford65,Ford87,Ford88}
\BEQ \label{3.26} 
\partial_t^2 s +\lambda \partial_t s + \omega^2 s = \zeta \;\; , \;\; \zeta = \frac{1}{m}\eta_p +\lambda \eta_s + \partial_t \eta_s 
\EEQ
The second moments of the `composite noise' $\zeta(t)$ are readily found, first in frequency space
\begin{align}
\big<\big\{ \wht{\zeta}(\nu), \wht{\zeta}(\nu') \big\}\big> 
= \frac{2\hbar\lambda}{m}\, \nu \coth\left(\frac{\hbar\nu}{2T}\right)\: \delta(\nu+\nu') \;\; , \;\; 
\big<\big[ \wht{\zeta}(\nu), \wht{\zeta}(\nu') \big]\big> &= -\frac{2\hbar\lambda}{m}\:\nu\,\delta(\nu+\nu')
\end{align}
and then for the times
\begin{subequations} \label{eq:zeta}
\begin{align} 
\left\langle \left\{ \zeta(t), \zeta(t') \right\}\right\rangle 
&=  \frac{2\hbar\lambda}{\pi m} \int_0^{\infty} \!\D\nu\: \nu \coth\left(\frac{\hbar\nu}{2T}\right)\cos(\nu(t-t'))
\:=:\: \frac{\hbar\lambda}{\pi m}\, I\left(\frac{\hbar}{2T}, t-t'\right) 
\label{eq:zeta-a}  \\
\left\langle \left[ \zeta(t), \zeta(t') \right]\right\rangle 
&= \frac{2\hbar\lambda}{\II \pi m} \int_0^{\infty} \!\D\nu\: \nu \sin(\nu(t-t')) 
\hspace{2.1truecm}\:=\:\: \frac{2\II\hbar\lambda}{m}\: \delta'\left(t-t'\right) 
\label{eq:zeta-c}
\end{align}
\end{subequations}
which {is the noise-averaged form of} the ones proposed by Ford {\em et al.} 
\cite[eqs. (2.2,2.3)]{Ford88}.\footnote{Eq.~(\ref{eq:zeta-a}) 
can also be derived directly from the requirement that the {\sc qfdt} holds \cite{Pasq84}.} 
Since the `composite noise' $\zeta$ contains derivatives,
it should be rougher than the individual noises $\eta_s,\eta_p$. 
This feature might be useful for the construction of numerical approximation schemes 
for the {\sc qle}s (\ref{1.2}).\footnote{Since the $\nu\to\infty$-behaviour in (\ref{3.25}) is less singular than in 
(\ref{eq:zeta-a}), it does not depend on an eventual cut-off, see \cite[sec. 3.3]{Gard04}.}

Finally, converting (\ref{3.25}) to real times leads to (as shown in appendix~B) 
\BEQ \label{3.29} 
\left\langle\bigl\{ \eta_s(t), {\eta}_p(t') \bigr\}\right\rangle 
= \lambda T \coth\left( \frac{\pi T}{\hbar}(t-t')\right) \;\; , \;\;
\left\langle\bigl[ {\eta}_s(t), {\eta}_p(t') \bigr]\right\rangle 
=\II\hbar\lambda \, \delta(t-t')
\EEQ
These are equivalent to the averages given in (\ref{1.5}).\footnote{In appendix~B, we shall study in
more detail the essential singularities of the average anti-commutator 
$\left\langle\left\{ \eta_s(t), {\eta}_p(t') \right\}\right\rangle 
=\frac{\hbar\lambda}{2\pi} J\left(\frac{\hbar\nu}{2T},t-t'\right)$, 
with $J(a,\tau) := \int_{\mathbb{R}}\!\D\nu\: \coth\left(a \nu\right) \sin\left(\nu\tau\right)$.}   
Hence, {\it the {\sc qfdt} implies non-markovian noise correlators}.

Our discussion of the stationary state has produced the necessary ingredients  
to undertake the study of the relaxation behaviour, starting from
an arbitrary initial state. For the specific example of the {\sc qho}, 
it is enough to take up the formal solution (\ref{2.7}) and to use
the quantum noise correlators (\ref{1.5}).  While the responses are (of course !) 
unchanged with respect to the treatment of section~2, the correlators
will now obey the {\sc qfdt}, by construction. The explicit results have been known since a long time, 
see \cite{Ford65,Ford87,Ford88,Gard04,Weis12} and references therein,
we shall not repeat the explicit calculation. 

\textcolor{blau}{\underline{\textcolor{black}{Summarising:}}} 
{\it if the only non-vanishing second moments of the noises $\eta_s,\eta_p$ are given by
(\ref{3.29}) (or equivalently by (\ref{3.25})), then the four criteria {\bf (A,B,C,D)} 
for a quantum dynamics of the harmonic oscillator (\ref{1.2}) are satisfied.}

\section{An example: the quantum spherical model}

The classical spherical model was initially formulated as an exactly solvable variant of the Ising model, 
in $d$ spatial dimensions \cite{Berl52,Lewi52}. For short-ranged interactions, and
dimensions $2<d<4$, its second-order phase transition is in an universality class distinct from mean-field theory. 
The quantum spherical model is defined by the hamiltonian \cite{Ober72,Henk84a,Vojt96,Oliv06,Wald15}
\BEQ \label{4.1}
H = \sum_{\vec{n}\in\mathscr{L}} \left( -J \sum_{j=1}^d S_{\vec{n}} S_{\vec{n}+\vec{e}_{j}} 
+ \frac{\mu}{2} S_{\vec{n}}^2 + \frac{g}{2} P_{\vec{n}}^2 \right) \;\; , \;\; 
\left[ S_{\vec{n}},P_{\vec{m}} \right] = \II \hbar\, \delta_{\vec{n},\vec{m}} 
\EEQ
where $S_{\vec{n}}$ is the spin operator at the site $\vec{n}\in\mathscr{L}\subset \mathbb{Z}^d$ of a hyper-cubic lattice 
$\mathscr{L}$ with $|\mathscr{L}|=\cal N$ sites, 
$P_{\vec{n}}$ is the canonically conjugate momentum, the constants $J$ and $g$ describe the nearest-neighbour interactions 
and the kinetic energy, respectively, and $\mu$ is the
Lagrange multiplier whose value is fixed by the (mean) spherical constraint 
$\left\langle \sum_{\vec{n}} S_{\vec{n}}^2 \right\rangle={\cal N}$. 
In momentum space,\footnote{Adapting eq.~(\ref{gl:Fourier}), we write  
$\wht{S}(t,\vec{q}):=(2\pi)^{-d/2}\int_{\mathbb{R}^d}\D\vec{r}\: \e^{-\II\vec{q}\cdot\vec{r}}S(t,\vec{r})$ etc.} 
the equations of motion of the different modes decouple, 
such that the quantum Langevin equations are 
\BEA
\partial_t \wht{S}(t,\vec{q}) &=& g \wht{P}(t,\vec{q}) + \wht{\eta}_s(t,\vec{q}) \nonumber \\
\partial_t \wht{P}(t,\vec{q}) &=& -\frac{1}{g} \left(\mathfrak{z}(t) +\omega(\vec{q}) \right)\wht{S}(t,\vec{q})
-\lambda\wht{P}(t,\vec{q})+\wht{\eta}_p(t,\vec{q})
\label{4.2}
\EEA
with $\omega(\vec{q}) = 2 \sum_{j=1}^d ( 1 - \cos q_j)$ and $\mu(t) = d + \mathfrak{z}(t)$. 
The spherical constraint, which in momentum space becomes 
$(2\pi)^{-d}\int_{\cal B}\D\vec{q}\: \left\langle\wht{S}(t,\vec{q})\wht{S}(t,-\vec{q})\right\rangle=1$,
maintains a coupling between the modes and fixes $\mathfrak{z}(t)$ self-consistently
(${\cal B}=[-\pi,\pi]^d$ is the Brillouin zone). {Each mode relaxes separately to its equilibrium state 
whose (semi-)classical or quantum nature is determined by the noises.} 

{Since the Lagrange multiplier $\mathfrak{z}(t)$ is time-dependent, this linear system {\em cannot} be solved
via a straightforward matrix exponentiation (the time-dependent coefficient matrices do not commute for different times). 
Rather, the quite technical Magnus expansion \cite{Magn54,Blan09} must be considered. However, the long-time
behaviour follows from a single first-order Langevin equation, which we shall now derive via a long-time scaling
limit on the Langevin equations (\ref{4.2}) and the associated noises. Considering this over-damped Langevin equations
considerably simplifies the explicit solutions of the dynamics, to be presented in future work.} 

{\bf 1.} First, we consider the Bedeaux-Mazur noise correlators (\ref{1.3}), enhanced by {\em spatial locality}. 
In momentum space, the noise correlators read 
\BEA
\left\langle \wht{\eta}_p(t,\vec{q}) \wht{\eta}_p(t',\vec{q}')\right\rangle 
&\!\!=& \!\! \frac{\lambda\hbar\sqrt{\mathfrak{z}(t)+\omega(\vec{q})}}{g} 
\coth\left(\frac{\hbar\sqrt{\mathfrak{z}(t)+\omega(\vec{q})}}{2T}\right) \delta(t-t') \delta(\vec{q}+\vec{q}') 
\nonumber \\
\left\langle \wht{\eta}_s(t,\vec{q}) \wht{\eta}_p(t',\vec{q}')\right\rangle &\!\!=& 
\!\!- \left\langle \wht{\eta}_p(t,\vec{q}) \wht{\eta}_s(t',\vec{q}')\right\rangle \:=\: 
\frac{1}{2}\II\hbar\lambda\, \delta(t-t') \delta(\vec{q}+\vec{q}') 
\label{4.3}
\EEA
Information on the long-time behaviour can be obtained by making the scaling transformation
\BEQ \label{gl:4.4-scal}
\tau := g t \;\; , \;\; 
\wht{\eta}_s(t,\vec{q}) = g^{3/2} \wit{\eta}_s(\tau,\vec{q}) \;\; , \;\; 
\wht{\eta}_p(t,\vec{q}) = g^{-1/2}\wit{\eta}_p(\tau,\vec{q}) \;\; , \;\; 
\wht{S}(t,\vec{q}) = g^{1/2}\wit{S}(\tau,\vec{q})
\EEQ
and then by considering the long-time scaling limit $g \to 0$, $t\to \infty$ such that $\tau=g t$ is kept fixed. 
Combining eqs.~(\ref{4.2},\ref{4.3}) leads to the equation of motion for the rescaled spin $\wit{S}(\tau,\vec{q})$ 
\BEQ \label{4.5}
\underbrace{g^2\partial_{\tau}^2 \wit{S}(\tau,\vec{q}) - g^2 \partial_{\tau}\wit{\eta}_s(\tau,\vec{q})}_{\mbox{\rm O($g^2$) $\to 0$}} = 
- \left(\mathfrak{z}(\tau) +\omega(\vec{q}) \right)\wit{S}(\tau,\vec{q})-\lambda g \partial_{\tau}\wit{S}(\tau,\vec{q})
+ \wit{\eta}(\tau,\vec{q}) 
\EEQ
with the new effective noise 
$\wit{\eta}(\tau,\vec{q}):=\sigma \wit{\eta}_s(\tau,\vec{q}) + g \wit{\eta}_p(\tau,\vec{q})$ 
and we also substituted $\mathfrak{z}(t)\mapsto \mathfrak{z}(\tau)$.  
The left-hand side of (\ref{4.5}) vanishes in the scaling limit such that a classical over-damped equation of motion remains.
In the over-damped case under consideration, $\lambda$ must be considered large, so that we actually let $\lambda\to\infty$, but such
that $\sigma := \lambda g$ is being kept fixed. Then the scaling limit (\ref{gl:4.4-scal}) is well-defined. We have the final 
semi-classical Langevin equation
\begin{subequations} \label{4.6}
\begin{align}
&\sigma \partial_{\tau}\wit{S}(\tau,\vec{q}) +  
         \bigl(\mathfrak{z}(\tau) +\omega(\vec{q}) \bigr)\wit{S}(\tau,\vec{q}) = \wit{\eta}(\tau,\vec{q}) 
\label{4.6a} \\
& \left\langle \wit{\eta}(\tau,\vec{q}) \wit{\eta}(\tau',\vec{q}')\right\rangle = 
\sigma \hbar \sqrt{\mathfrak{z}(\tau)+\omega(\vec{q})\,}\, 
\coth\left(\frac{\hbar\sqrt{\mathfrak{z}(\tau)+\omega(\vec{q})}}{2T}\right) \delta(\tau-\tau') \delta(\vec{q}+\vec{q}') 
\label{4.6b}
\end{align}
\end{subequations}
{along with $\left\langle \bigl[ \wit{\eta}(\tau,q),\wit{\eta}(\tau',q')\bigr]\right\rangle=0$.} 
In eq.~(\ref{4.6}), two limit cases can be recognised: 
\begin{enumerate}
\item[(a)] \underline{$T\to\infty$}. Eq.~(\ref{4.6b}) reduces to white noise 
$\left\langle \wht{\eta}(\tau,\vec{q}) \wht{\eta}(\tau',\vec{q}')\right\rangle = 2T \sigma \delta(\tau-\tau')\delta(\vec{q}+\vec{q}')$
and eq.~(\ref{4.6a}) describes the relaxation towards classical thermal equilibrium.  
Unsurprisingly, for sufficiently large temperatures, one is back to an effectively classical system. 

This appears analogous to the semi-classical behaviour of the quantum spherical model with Lindblad dynamics \cite{Wald18}. 
\item[(b)] \underline{$T\to 0$}. Then the noise correlator (\ref{4.6b}) reduces to 
\BEQ
\left\langle \wht{\eta}(\tau,\vec{q}) \wht{\eta}(\tau',\vec{q}')\right\rangle = 
\sigma\hbar \sqrt{\mathfrak{z}(\tau)+\omega(\vec{q})\,}\, \delta(\tau-\tau') \delta(\vec{q}+\vec{q}') 
\EEQ
Therefore, at vanishing temperature, the model shows some new type of non-classical behaviour, 
but which comes from a markovian dynamics. Further details will be presented elsewhere.  
\end{enumerate}

{{\bf 2.} We now present the same analysis for the quantum noises (\ref{1.5}) derived in section~3. The only
non-vanishing noise anti-commutators and commutators are
\BEQ
\left\langle\bigl\{ \wht{\eta}_s(t,\vec{q}) , \wht{\eta}_p(t',\vec{q}') \bigr\}\right\rangle 
= \frac{\hbar\lambda}{2\pi} J\left(\frac{\hbar}{2T}, t-t'\right) \delta(\vec{q}+\vec{q}') 
\;\; , \;\; 
\left\langle\bigl[ \wht{\eta}_s(t,\vec{q}) , \wht{\eta}_p(t',\vec{q}') \bigr]\right\rangle 
=\II\hbar\lambda \delta(t-t') \delta(\vec{q}+\vec{q}')
\EEQ
where the function $J(a,\tau)$ is analysed in appendix~B. }
{We apply the scaling transformation 
\BEQ \label{gl:4.8-scal}
\tau := g t \;\; , \;\; 
\wht{\eta}_s(t,\vec{q}) = g^{0} \wit{\eta}_s(\tau,\vec{q}) \;\; , \;\; 
\wht{\eta}_p(t,\vec{q}) = g^{0}\wit{\eta}_p(\tau,\vec{q}) \;\; , \;\; 
\wht{S}(t,\vec{q}) = g^{1}\wit{S}(\tau,\vec{q})
\EEQ
to eqs.~(\ref{4.2}), re-written in the form 
\BEQ
\underbrace{g^2 \partial_{\tau} \wit{S}(\tau,\vec{q})}_{\mbox{\rm O$(g^2)\to 0$}} 
= - \bigl(\mathfrak{z}(\tau) + \omega(\vec{q})\bigr)\wit{S}(\tau,\vec{q}) 
- \sigma \partial_{\tau}\wit{S}(\tau,\vec{q})  + \wit{\zeta}(\tau,\vec{q}) + \wit{\xi}(\tau,\vec{q})
\EEQ
with the two auxiliary noises  
\BEQ \label{gl:4.11-bruit}
\wit{\zeta}(\tau,\vec{q}) :=  \wit{\eta}_p(\tau,\vec{q}) + \sigma g^{-2} \wit{\eta}_s(\tau,\vec{q})
\;\; , \;\; 
\wit{\xi}(\tau,\vec{q}) :=  \partial_{\tau}\wit{\eta}_s(\tau,\vec{q})
\EEQ
and consider the scaling limit $t\to\infty$, $g\to 0$ such that $\tau=g t$ is kept fixed.\footnote{Although the second term in the
noise $\wit{\zeta}$ in  (\ref{gl:4.11-bruit}) diverges, it does {\em not} contribute, neither to the noise commutators, nor to the
noise anti-commutators.} Herein, the rescaled dissipation rate $\sigma$ and the rescaled temperature $\Theta$ are
\BEQ
\sigma = \lambda g \;\; , \;\; \Theta = T g^{-1}
\EEQ
Then the noise commutator
\BEQ
\left\langle\bigl[ \wit{\zeta}(\tau,\vec{q}) + \wit{\xi}(\tau,\vec{q}),\wit{\zeta}(\tau',\vec{q}') 
+ \wit{\xi}(\tau',\vec{q}')\bigr\}\right\rangle =
2\II\hbar\sigma \delta'(\tau-\tau') \delta(\vec{q}+\vec{q}') 
\EEQ
reproduces (\ref{eq:zeta-c}) and noise anti-commutator, with $I(a,\tau)=\partial_{\tau}J(a,\tau)$
\BEQ
\left\langle\bigl\{ \wit{\zeta}(\tau,\vec{q}) + \wit{\xi}(\tau,q),
\wit{\zeta}(\tau',\vec{q}') + \wit{\xi}(\tau',\vec{q}')\bigr\}\right\rangle =
\frac{\hbar\sigma}{\pi} I\left(\frac{\hbar}{2\Theta},\tau-\tau'\right) \delta(\vec{q}+\vec{q}') 
\EEQ
reduces exactly to the form (\ref{eq:zeta-a}), with the mass set to $m=1$ and up to spatial correlations. 
Therefore, instead of (\ref{4.6}), the Langevin equations  now become
\begin{subequations} \label{4.15}
\begin{align}
\sigma \partial_{\tau}\wit{S}(\tau,\vec{q}) +  
         \bigl(\mathfrak{z}(\tau) +\omega(\vec{q}) \bigr)\wit{S}(\tau,\vec{q}) &= \wit{\eta}(\tau,\vec{q}) 
\label{4.15a} \\
\left\langle \bigl\{ \wit{\eta}(\tau,\vec{q}), \wit{\eta}(\tau',\vec{q}')\bigr\}\right\rangle &= 
\frac{\sigma \hbar}{\pi} I\left(\frac{\hbar}{2\Theta},\tau-\tau'\right) \delta(\vec{q}+\vec{q}') 
\label{4.15b} \\ 
\left\langle \bigl[ \wit{\eta}(\tau,\vec{q}), \wit{\eta}(\tau',\vec{q}')\bigr]\right\rangle &= 
2\II\sigma \hbar\, \delta'(\tau-\tau') \delta(\vec{q}+\vec{q}')
\label{4.15c}
\end{align}
\end{subequations}
Clearly, we have two limit cases, using the asymptotics of $I(a,\tau)$ from appendix~B:
\begin{enumerate}
\item[(a)] \underline{$\Theta\to\infty$}. One is back to classical white noise. 
\item[(b)] \underline{$\Theta\to 0$}. The eq.~(\ref{4.15b}) becomes a pure quantum noise, with memory 
\BEQ
\left\langle \bigl\{ \wit{\eta}(\tau,\vec{q}), \wit{\eta}(\tau',\vec{q}')\bigr\}\right\rangle 
= -\frac{2\sigma \hbar}{\pi}  \left(\frac{1}{\tau - \tau'}\right)^{2} \delta(\vec{q}+\vec{q}')
\EEQ
and eq.~(\ref{4.15c}) is kept. 
\end{enumerate}
%
\begin{table}\begin{center}
\begin{tabular}{|c|ll|} \hline
                           & Bedeaux \& Mazur       & quantum  \\ \hline
dissipation $\lambda$      & increases for $g\to 0$ & increases for $g\to 0$\\
temperature $T$            & constant               & decreases for $g\to 0$ \\
line $T=0$                 & neutral                & unstable \\
noises $\eta(t), \eta(t')$ & commute for $t\ne t'$  & do not commute \\ \hline
\end{tabular}
\end{center}
\caption[tab1]{Some physical characteristics of the scaling transformations 
(\ref{gl:4.4-scal},\ref{gl:4.8-scal}), respectively, leading for $g\to 0$
to the overdamped Langevin equations (\ref{4.6a},\ref{4.15a}), depending on the chosen dynamics. \label{tab1}}
\end{table}
%
In table~\ref{tab1}, we collect some characteristics of the scaling transformation which lead to the overdamped
Langevin equation (\ref{4.6a},\ref{4.15a}), for either Bedeaux-Mazur noise (\ref{1.3}) or else the quantum noise (\ref{1.5}). 
The scaling limit generically leads to the overdamped limit $\lambda\to\infty$. In the quantum case, the zero-temperature
line $T=0$ is unstable which means that the long-time behaviour for all $T>0$ should be the same as for the classical limit
$T\to\infty$, while for Bedeaux-Mazur noise $T$ is not rescaled. Another important distinction concerns whether the
noise commutator $\left\langle \bigl[ \eta(t), \eta(t')\bigr]\right\rangle$ vanishes for different times $t,t'$ or not. 
}

{\bf 3.} Finally, we illustrate some aspects of the non-markovian noise correlator (\ref{3.29}). 
Consider a single {\sc qho}, described by the
{\sc qle} (\ref{1.2}), but with the noises (\ref{1.5}). Following the lines of the treatment of section~2, 
the equal-time spin-spin correlator $C_{+}^{(s)}(t):=C_{+}^{(s)}(t,t)$ now becomes
\BEQ \label{gl:corr-t}
C_+^{(s)}(t) = \frac{T}{2m}
\int_0^t \!\D\tau \int_0^t \!\D\tau'\: \coth\left(\frac{\pi T}{\hbar} (\tau-\tau')\right) 
\left[ \e^{-\Lambda_+(t-\tau) - \Lambda_-(t-\tau')} - \e^{-\Lambda_-(t-\tau) - \Lambda_+(t-\tau')} \right]
\EEQ
where $\Lambda_{\pm}$ are given in (\ref{2.2}). For notational simplicity, and analogously to section~2, 
the initial conditions were chosen such that the non-stationary terms vanish.\footnote{{The first-order equations
(\ref{1.2},\ref{1.5}) readily lead to a convergent integral representation of the correlator $C_{+}^{(s)}(t)$ without
the need of an explicit regularisation.}}

\begin{figure}[tb]
\begin{center}
\includegraphics[width=\textwidth]{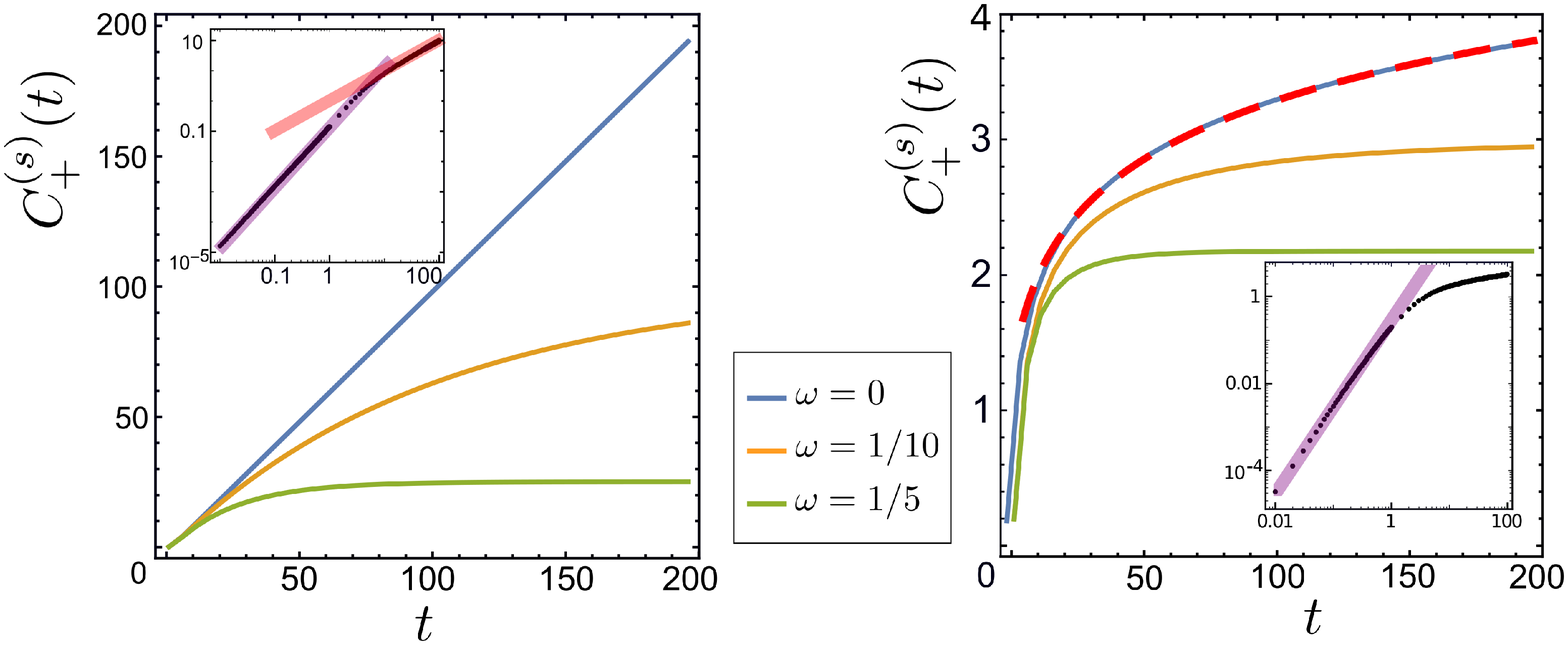} 
\end{center}
\caption[fig3]{Time-dependence of the spin-spin correlator $C_+^{(s)}(t)$ (\ref{gl:corr-t}) of the {\sc qho}, 
for $m=\lambda=1$ and $\omega = [0,\frac{1}{10}, \frac{1}{5}]$ from top to bottom. \\
(a) \underline{\bf Left panel}: $T=1$. At large times, the correlator saturates for $\omega > 0$, 
while it diverges linearly for $\omega = 0$. 
The inset shows, for $\omega = 0$, the early-time behaviour $C_+^{(s)}\sim t^2$. 
The violet and red curves correspond to $\sim t^2$ and $\sim t$ behaviour, respectively and are guides to the eye. \\
(b) \underline{\bf Right panel}: $T=0$. The correlator saturates for $\omega \neq0$ but grows logarithmically for $\omega = 0$ 
(the red-dashed line is a guide to the eye). The inset shows again the short-time behaviour for $\omega=0$ 
and the violet curve $\sim t^2$ is a guide to the eye. 
\label{fig3} }
\end{figure}

In figure~\ref{fig3}, we illustrate the time-dependence of $C_+^{(s)}(t)$, 
for several values of $\omega$ and for two values of $T$ (units are such that $\hbar=1$). 
\begin{enumerate}
\item[(A)] The case $\omega>0$ corresponds to a harmonic coupling. In this case, for both $T=0$ and $T=1$, the correlator saturates
rapidly, although the absolute scales are different. On the other hand, 
if one were to consider the {\sc qle} (\ref{1.2}) of the harmonic oscillator 
with a classical white noise instead, from (\ref{2.34}) we would have found the constant $C_+^{(s)}=T/(m \omega^2)$. 
The values of that constant indeed agree with the plateau values of the left panel in figure~\ref{fig3}. 
Hence, for a finite temperature $T>0$, the effect of the quantum correlator (\ref{1.5}) disappears after a finite cross-over time 
and for sufficiently large times, the system effectively behaves as if it were classical. 
For $T=0$, however, the values of the stationary correlator $C_+^{(s)}(\infty)$ are distinct from the classical values and the
saturation occurs much faster.  
\item[(B)] The case $\omega=0$ can either be interpreted as a critical mode of a spin system or else as a diffusing quantum particle. 
In the latter case, $C_+^{(s)}(t) = \left\langle r^2(t)\right\rangle$ can be interpreted as the variance of the displacement 
of the quantum particle. Here, we find an important qualitative difference as a function of $T$. For $T=1$, 
the long-time variance $C_+^{(s)}(t)\sim t$ as for a classical particle (left panel), whereas 
for the  case $T=0$ a logarithmic growth $C_+^{(s)}(t)\sim \ln t$ typical for quantum diffusion \cite{Haki85} is found (right panel). 

In the insets, for $\omega=0$ the early-time behaviour $C_+^{(s)}(t)\sim t^{2/z}$ is further illustrated. 
For both $T=0$ and $T=1$, an effectively ballistic behaviour with dynamical exponent $z=1$ is seen. 
This persists until the particles begin to encounter the obstacles created by the
interactions with the bath. For $T=1$, one finds the cross-over to classical behaviour 
with $z=2$ while for $T=0$ one crosses over to
quantum diffusion where $C_+^{(s)}(t)\sim \ln t$ grows logarithmically \cite{Haki85} and effectively $z=\infty$. 
\end{enumerate}

Turning now to the quantum spherical model, one has the {\sc qle} eqs.~(\ref{4.2}), 
with the noise correlators (\ref{3.29}) for each mode, but with 
an extra factor $\delta(\vec{q}+\vec{q}')$. These modes are coupled by the spherical constraint, which 
in momentum space can be formally written as
\BEQ \label{eq:contrainte-q}
\int_{\cal B} \frac{\!\D\vec{q}}{(2\pi)^d}\: \wht{C}_+^{(s)}(t,t;\vec{q},-\vec{q}) = 1 \ .
\EEQ
This makes the the spherical parameter $\mathfrak{z}=\mathfrak{z}(t)$ time-dependent. 
As a consequence, the matrix equation (\ref{4.2}) {\em cannot} be solved by a formal matrix exponentiation. 
In principle, more elaborate tools such as a Magnus expansion or scaling ans\"atze should be tried
but an explicit solution of (\ref{4.2}) is still unknown. If such a solution could be found, 
solving eq.~(\ref{eq:contrainte-q}) would determine the Lagrange multiplier $\mathfrak{z}(t)$. 
Since the form of this equation is likely to be very
different from the Volterra integral equations found for the classical spherical model \cite{Ronc78,Godr00b}, 
its solution $\mathfrak{z}(t)$ should also be different from the classical result. 
We hope to return to this difficult open problem elsewhere.  

Although the response functions do not depend explicitly on the noises, 
they do contain the Lagrange multiplier $\mathfrak{z}(t)$. 
Therefore, via the spherical constraint ({\ref{eq:contrainte-q}), the quantum noises will influence the long-time
behaviour and the dynamical scaling of the two-time responses $R^{(s,p)}(t,t')$.

\section{{\it Concludendum est}}

We presented attempts for an axiomatic construction of quantum Langevin equations. These have been based on four physical requirements 
which appeared reasonable to us, namely {\bf (A)} canonical equal-time commutators, 
{\bf (B)} the Kubo formul{\ae} for spin and momentum, {\bf (C)} the virial theorem and {\bf (D)} the
quantum fluctuation-dissipation theorem. The central role of the {\sc qfdt} in this discussion stems 
{not only from its recognised fundamental importance in quantum statistical mechanics \cite{Hang05,Weis12,Ford17}} 
but also from its being an immediate consequence of an important dynamical symmetry for quantum equilibrium states \cite{Sieb15,Aron18}. 
We used the harmonic oscillator equations of motion (\ref{1.2}) as a scaffold to derive the second moments of
the two noises $\eta_s,\eta_p$. This form was motivated by the study of Bedeaux and Mazur \cite{Bede01,Bede02} 
although it turned out that their specific proposal (\ref{1.3}), which is markovian, 
should be seen as a semi-classical description, 
since it only satisfies the criteria {\bf (A,B,C)} while the condition {\bf (D)} can only
be satisfied in the classical limit $\hbar\to 0$. 
Furthermore, their noise correlators do depend on the specific model parameters $m,\omega$ of the harmonic
oscillator and there is no obvious generalisation beyond this specific model. 
On the other hand, not requiring the Markov property from the outset  
does lead to a different approach where  the noise correlators (\ref{1.5}) (i) are model-independent
and only contain bath properties (e.g. the temperature $T$) and the ohmic dissipation constant $\lambda$ 
(ii) are non-markovian as required from the validity
of the {\em quantum} fluctuation-dissipation theorem for all $\hbar>0$. 
The noise correlators (\ref{1.5}) are equivalent to the well-known second-order quantum Langevin equation of {\sc fkm}, of 
a particle in an external potential \cite{Ford65,Ford88,Weis12}. 
It follows that the quantum noise correlators can indeed be derived from
an Caldeira-Leggett-type independent oscillator model of the bath \cite{Ford65,Ford87,Ford88,Gard04,Weis12}. 
{Still, the present re-derivation of a well-established result is of interest 
since the implicit assumptions behind each derivation are not entirely identical: 
while the derivation according to {\sc fkm} \cite{Ford65,Ford87,Ford88} considers a particle in an arbitrary potential $V(s)$ 
and assumes explicitly a bath made from harmonic oscillators, 
our approach focusses on a single harmonic oscillator and proceeds to specify the second moments
of the noise correlators, but does not assume neither gaussianity nor anything else specific about the bath. 
These two approaches produce the noise correlators (\ref{1.5},\ref{eq:zeta}) which are mathematically equivalent.  
It follows that the respective auxiliary assumptions are not really required. 
Hence the noise correlators (\ref{1.5}) should be generally valid for the description of quantum dynamics, 
both for arbitrary potentials $V(s)$ as well as for a bath with non-harmonic degrees of freedom. In addition, the four properties
{\bf (A,B,C,D)} should be viewed as built-in features of Caldeira-Legget-type oscillator models of the bath.} 
The present formalism also makes the symplectic structure more explicit. 

On the other hand, different choices for the noise correlators are possible. Indeed, their generic form can be described
in terms of two anti-symmetric functions $\chi(t)$ and $\psi(t)$ (see appendix~C) 
and two symmetric functions $\alpha(t)$ and $\beta(t)$ (see section~3.4), but
these functions do contain the specific model parameters $m,\omega$. 
All four criteria {\bf (A,B,C,D)} are satisfied by these solutions but whose applicability is restricted to the harmonic oscillator. 
Eq.~(\ref{1.5}) is merely the only solution which does not depend on these parameters. Additional physical criteria, beyond
considerations of maximal simplicity, are required to further specify these functions left undetermined and not contained
in the Ford-Kac-Mazur construction \cite{Ford65,Ford87,Ford88}.

We also used the quantum spherical model as an illustration how to set up the quantum equations of motion for 
an analysis of the long-time relaxational properties of this many-body problem. It turns
out that such a {\em constrained quantum dynamics} leads to a formidable problem which 
has obtained a lot of attention, already since the early days of Dirac \cite{Dirac58} 
{and has been actively discussed recently, 
see e.g. \cite{Timp18,Wald16,Wald18,Gusta} and references therein. We presented a simple scaling argument which focusses on the
long-time limit where an effective constrained Langevin equation should become more easily treatable. 
Work along these lines is in progress, with the aim of studying non-mean-field many-body quantum dynamics in higher spatial dimensions.}

\noindent
{\bf Acknowledgements:}
We are grateful to A. Chiocchetta, M. D. Coutinho-Filho, A. Gambassi, G.T. Landi, H. Spohn and L. Turban for useful discussions. 
SW thanks the LPCT Nancy for warm hospitality, where part of this work was carried out.
RA acknowledges financial support from CAPES (Funda\c{c}\~ao do Minist\'erio Brasileiro da Educa\c{c}\~ao), 
process number 88881.132083/2016-01.


\newpage


\appsection{A}{}

We recall known results on three formul{\ae} from quantum statistical mechanics. The notation will be adapted
such as to be in agreement with the definition (\ref{gl:Fourier}) of the Fourier transform. 

\subsection{Kubo formula}

The Kubo formul{\ae} are a generic result of linear-response theory, even outside a stationary state.
For a quantum-mechanical system, its well-known time-dependent form is \cite{Cugl03}
\BEQ \label{gl:A:Kubo}
R(t-t') = \frac{2\II}{\hbar}\, \Theta(t-t') C_{-}(t,t') 
= \frac{2\II}{\hbar}\, \Theta(t-t') \demi \left\langle \left[ s(t), s(t') \right]\right\rangle
\EEQ
where for the sake of notational simplicity, 
we restrict attention to the response of a magnetic moment with respect to its canonically conjugated magnetic field. 

In frequency-space, and for a stationary state, the commutator is written as follows
\BEQ \label{gl:A:2} 
\demi \left\langle \left[ \wht{s}(\nu), \wht{s}(\nu') \right]\right\rangle 
=: \wht{C}_{-}(\nu,\nu') =: \delta(\nu+\nu')\, \wht{C}_{-}(\nu)
\EEQ
Using (\ref{gl:Fourier}), the Kubo formula (\ref{gl:A:Kubo}) becomes,  if only $t-t'>0$ 
\BEA
\frac{1}{\sqrt{2\pi\,}\,}\int_{\mathbb{R}} \!\D\nu\: \e^{\II\nu(t-t')} \wht{R}(\nu) 
&\stackrel{.}{=}& \frac{2\II}{\hbar} \frac{1}{2\pi} \int_{\mathbb{R}^2} \!\D\nu \D\nu'\: \e^{\II\nu t + \II\nu' t'} 
\demi  \left\langle \left[ \wht{s}(\nu), \wht{s}(\nu') \right]\right\rangle \nonumber \\
&=& \frac{2\II}{\hbar} \frac{1}{2\pi} \int_{\mathbb{R}} \!\D\nu\: \e^{\II\nu t + \II\nu' t'} 
    \delta(\nu+\nu') \wht{C}_{-}(\nu) \nonumber \\
&=& \frac{2\II}{\hbar} \frac{1}{2\pi} \int_{\mathbb{R}} \!\D\nu\: \e^{\II\nu (t-t')} \wht{C}_{-}(\nu)
\EEA
where (\ref{gl:A:2}) was used in the second line. Since this holds for all $t-t'>0$, we have the relation between the amplitudes 
\BEQ \label{gl:A:Kubo-Fourier} 
\wht{R}(\nu) \stackrel{.}{=} -\frac{2}{\hbar\II} \frac{1}{\sqrt{2\pi\,}\,}\: \wht{C}_{-}(\nu)
\EEQ
which we shall need below in the derivation of the {\sc qfdt}.  Momentum responses are treated analogously. 

\subsection{Virial theorem}

The virial theorem is a well-known result of equilibrium statistical mechanics, with many important applications \cite{Harw06,Schn08}. 
A first quantum-mechanical derivation was given by Fock \cite{Fock30}. 
Consider a system of $N$ interacting particles with masses $m_n$, and with the hamiltonian 
\BEQ
H = \sum_n \left( \frac{p_n^2}{2m_n} + V(\{ x_n\}) \right)
\EEQ
Use the momentum operator $p_n = \frac{\hbar}{\II} \frac{\D}{\D x_n}$ and define $\mathscr{V} := \sum_n x_n p_n$. This gives 
\BEQ
\frac{\D \mathscr{V}}{\D t} = \frac{\II}{\hbar} \left[ H, \mathscr{V} \right] = 2 T - \sum_n x_n \frac{\D V}{\D x_n} 
\EEQ
where $T = \demi \sum_n p_n^2/m_n$ denotes the kinetic energy. 
In a stationary state, with the quantum-mechanical average $\langle\cdot\rangle$, 
one has $\left\langle \frac{\D \mathscr{V}}{\D t}\right\rangle=0$, hence
$2\left\langle T\right\rangle = \left\langle \sum_n x_n \frac{\D V}{\D x_n}\right\rangle$. 
Furthermore, if the potential is harmonic of degree $\alpha$ in all its arguments, 
that is $V(\lambda x_n) = \lambda^{\alpha} V(x_n)$ with $\lambda>0$ 
and for all $x_n$ with $n=1,\ldots,N$, this implies the {\em virial theorem}
\BEQ \label{gla:virial}
\left\langle T \right\rangle = \frac{\alpha}{2} \left\langle V \right\rangle
\EEQ
Such stationary averages exist for $\alpha>-2$. For the harmonic oscillator, $\alpha=2$, hence 
$\left\langle T \right\rangle =\left\langle V \right\rangle$. If the system is a single harmonic oscillator, with
$H = \frac{p^2}{2m} + \frac{m \omega^2}{2} x^2$, eq.~(\ref{gla:virial}) becomes
$\left\langle p^2\right\rangle = m^2 \omega^2 \left\langle x^2\right\rangle$,
or alternatively $C_{+}^{(p)}(t,t)=m^2\omega^2 C_{+}^{(s)}(t,t)$ as quoted  in the main text. 

For extensions to relativistic quantum mechanics, see \cite{Fock30}.

\subsection{Kubo-Martin-Schwinger relation and the {\sc qfdt}} 

{Following \cite{Cugl03}, we recall the derivation of the Kubo-Martin-Schwinger ({\sc kms}) relation for two-time quantities and
use it to obtain the quantum fluctuation-dissipation theorem ({\sc qfdt}).} 

{At equilibrium, quantum correlators between the time-dependent observables $A,B$ are defined as 
\BEQ
C_{AB}(t,t')=\langle A(t) B(t')\rangle := \frac{1}{Z} \Tr{A(t) B(t') \rho(0)}
\EEQ
where $\rho(0)=\exp\left( -H/T\right)$ is the density matrix and $Z:= \Tr{\rho(0)}$. In the Heisenberg representation,
$A(t) = e^{\II/\hbar\, Ht} A(0) e^{-\II/\hbar\, Ht}$. Then the quantum correlator can be re-expressed as follows
\BEA
\lefteqn{{C_{AB}(t,t') =} \;\frac{1}{Z} \Tr{B(t') \: e^{-H/T} A(t)}} \nonumber \\
&=& \frac{1}{Z} \Tr{ B(t') \:{\underbrace{\textcolor{black}{e^{-H/T} e^{\II/\hbar \,H t}}}} 
            \hspace{0.3truecm}A(0) \hspace{1.7truecm}e^{-\II/\hbar\, H t} \;\;} \nonumber \\
&=& \frac{1}{Z} \Tr{B(t') 
            \textcolor{black}{\underbrace{\textcolor{black}{\:e^{\II/\hbar\, H (t- \hbar/(\II T))} A(0)\:} 
            \textcolor{white}{e^{-\II/\hbar\, H (t- \hbar/(\II T))}}}} 
\hspace{-2.4truecm}{\overbrace{\textcolor{black}{e^{-\II/\hbar\, H (t- \hbar/(\II T))} 
\:e^{\II/\hbar\, H (- \hbar/(\II T))}}}} \: } \nonumber \\
&=& \frac{1}{Z} \Tr{B(t')\hspace{1.6truecm}A\left(t+\frac{\II\hbar}{T}\right)\hspace{2.5truecm} e^{-H/T} \; } 
\nonumber\\
&{=}& {C_{BA}\left(t',t+\frac{\II\hbar}{T}\right)} \label{kms1} 
\EEA
Symmetrising with respect to the times, the Kubo-Martin-Schwinger  relation (\ref{kms1}) becomes
\BEQ \label{kms2} 
C_{AB}(t-\frac{\II\hbar}{2T}, t') = C_{BA}(t',t+\frac{\II\hbar}{2T})
\EEQ
}
{In order to obtain the {\sc qfdt},} we specialise to the case when $A=B$ and consider
\BEQ
C_{AA}(t,t') = \demi \left\langle \left\{ A(t), A(t') \right\}\right\rangle 
               + \demi \left\langle \left[ A(t), A(t') \right]\right\rangle
=: C_{+}(t,t') + C_{-}(t,t')
\EEQ
Specifically, we shall take $A=s$ or $A=p$ in the text. 
Herein, $C_+(t,t')=C_+(t',t)$ is symmetric and $C_-(t,t')=-C_-(t',t)$ is antisymmetric. Then the {\sc kms} relation (\ref{kms2}) is
\BEQ
C_+\left(t',t+\frac{\II\hbar}{2T}\right) + C_-\left(t',t+\frac{\II\hbar}{2T}\right) 
= C_+\left(t-\frac{\II\hbar}{2T},t'\right) + C_-\left(t-\frac{\II\hbar}{2T},t'\right)
\EEQ
We use the symmetry and antisymmetry to recast this as 
\BD
C_+\left(t+\frac{\II\hbar}{2T},t'\right) - C_+\left(t-\frac{\II\hbar}{2T},t'\right)  
=  C_-\left(t+\frac{\II\hbar}{2T},t'\right) + C_-\left(t-\frac{\II\hbar}{2T},t'\right)
\ED
Next, use the Kubo formula (\ref{gl:A:Kubo}) and the stationarity of $C_+=C_+(t-t')$ and $R=R(t-t')$ at equilibrium. This gives
\BD
C_+\left(t-t'+\frac{\II\hbar}{2T}\right) - C_+\left(t-t'-\frac{\II\hbar}{2T}\right)  
=  -\frac{\II\hbar}{2} \left(  R\left(t-t'+\frac{\II\hbar}{2T}\right) + R\left(t-t'-\frac{\II\hbar}{2T}\right)\right)
\ED
Fourier-transforming with respect to $\tau=t-t'$, via (\ref{gl:Fourier}), produces the {\sc qfdt} 
\BEQ \label{gl:A:10}
\tanh\left( \frac{\hbar \nu}{2 T} \right) \wht{C}_{+}(\nu) = \frac{\II\hbar}{2}\, \wht{R}(\nu) 
\EEQ
Finally, using the Kubo formula (\ref{gl:A:Kubo-Fourier}) brings the {\sc qfdt} into the form
\BEQ \label{gl:A:qfdt} 
\frac{\wht{C}_{+}(\nu)}{\wht{C}_{-}(\nu)} = - \frac{1}{\sqrt{2\pi\,}\,} \coth\frac{\hbar\nu}{2T}
\EEQ
which is required in the text. Eq.~(\ref{gl:A:qfdt}) shows that 
$\wht{C}_{+}^{(s)}/\wht{C}_{-}^{(s)}=\wht{C}_{+}^{(p)}/\wht{C}_{-}^{(p)}$ 
is independent of the observable and only contains properties of the bath, as it should be. 

\appsection{B}{Computation of an integral}

We calculate the inverse Fourier transformation of (\ref{3.25}) and derive (\ref{3.29}). 
{Then the subtle point of singular contributions to these inverse Fourier transforms is treated.} 

\subsection{Regular functions}
Consider
\BEA
\lefteqn{\left\langle \left\{ \eta_s(t), \eta_p(t') \right\}\right\rangle 
\:=\: \frac{1}{2\pi} \int_{\mathbb{R}^2} \!\D\nu\D\nu'\: \e^{\II\nu t+\II\nu' t'} 
\left\langle\left\{\wht{\eta}_s(\nu),\wht{\eta}_p(\nu')\right\}\right\rangle}
\nonumber \\
&=& \frac{1}{2\pi} \int_{\mathbb{R}^2} \!\D\nu\D\nu'\: \e^{\II\nu t+\II\nu' t'}\, \frac{\hbar\lambda}{\II} 
\coth\left(\frac{\hbar\nu}{2T}\right)\, \delta(\nu+\nu') \nonumber \\
&=& \frac{\hbar\lambda}{2\pi \II} \int_{\mathbb{R}} \!\D\nu\: \e^{\II\nu(t-t')}\, 
\left[ \coth\left(\frac{\hbar\nu}{2T}\right) - \sign \nu + \sign \nu\right] 
\nonumber \\
&=& \frac{\hbar\lambda}{2\pi \II} \int_{\mathbb{R}} \!\D\nu\: \e^{\II\nu(t-t')}\,\sign \nu 
+ \frac{\hbar\lambda}{2\pi}\, 2 \int_{0}^{\infty} \!\D\nu\: \sin\left(\nu(t-t')\right) 
\left[ \coth\left(\frac{\hbar\nu}{2T}\right) - \sign \nu \right] 
\label{gl:B1}
\EEA
where (\ref{3.25}) was used in the second line. 
In the last line, we separate into two contributions: (i) a singular integral which must be interpreted as a distribution and 
(ii) a regular convergent term, where the antisymmetry of the integrand was also used when 
retaining only the principal value of that second integral.
This second term is given by the identity 
\cite[eq. (2.5.46.17)]{Prud1}\footnote{The entry is labelled incorrectly (2.5.46.7) in \cite{Prud1}. 
It corrects errors in \cite[(17.33.26)]{Grad07} and \cite[(2.9.3)]{Erde54}.}
\BEQ
\int_0^{\infty} \!\D x\: \sin(bx) \left[ \coth (ax) - 1 \right] = \frac{\pi}{2a} \coth\left(\frac{\pi b}{2a}\right) - \frac{1}{b}
\EEQ
The first term in (\ref{gl:B1}) is evaluated as a distribution \cite[p. 173]{Gelf64}: 
$\int_{\mathbb{R}}\!\D x\: \e^{\II\sigma x}\,\sign x = 2\II \sigma^{-1}$. 
We then obtain 
\BEQ \label{gl:B3}
\left\langle \left\{ \eta_s(t), \eta_p(t') \right\}\right\rangle 
= \underbrace{\frac{\hbar\lambda}{\pi} \frac{1}{t-t'} - \frac{\hbar\lambda}{\pi} \frac{1}{t-t'} }_{=0}
+ \lambda T \coth \left(\frac{\pi T}{\hbar} (t-t')\right)
\EEQ
which is (\ref{3.29}) in the main text. 
We point out that the singular terms, arising in the intermediate states of the calculation, cancel in the final result. 

Alternatively, the result (\ref{gl:B3}) can be obtained from the identity \cite[(4.3.91)]{Abra65} 
\BEQ \label{gl:B4}
\coth \frac{\hbar\nu}{2T} = \lim_{\vep\to 0^{+}} \frac{2T\nu}{\hbar} 
\left( \frac{1}{(\nu+\II\vep)(\nu-\II\vep)} + 2 \sum_{n=1}^{\infty} \frac{1}{(\nu+\II\nu_n)(\nu-\II\nu_n)} \right)
\;\; , \;\; \nu_n = \frac{2\pi T}{\hbar}\, n
\EEQ
Then, for $\tau>0$ 
\BEA
\lefteqn{\left\langle \left\{ \eta_s(\tau), \eta_p(0) \right\}\right\rangle 
\:=\: \frac{\hbar\lambda}{2\pi \II} \int_{\mathbb{R}} \!\D\nu\: \e^{\II\nu \tau}\, \coth\left(\frac{\hbar\nu}{2T}\right)} \nonumber \\
&=& \frac{\lambda T}{\II \pi} \lim_{\vep\to 0^{+}} \oint_{{\cal C}} \!\D\nu \: \left[ 
\frac{\nu \e^{\II\nu\tau}}{(\nu+\II\vep)(\nu-\II\vep)} 
+ \sum_{n=1}^{\infty} \frac{2 \nu\e^{\II\nu\tau}}{(\nu+\II\nu_n)(\nu-\II\nu_n)} \right] \nonumber \\
&=& \frac{\lambda T}{\II \pi}\, 2\pi\II \lim_{\vep\to 0^{+}} \left[ 
\frac{\II \vep \e^{-\vep\tau}}{2\II\vep} + \sum_{n=1}^{\infty} \frac{2\II\nu_n \e^{-\nu_n \tau}}{2\II\nu_n} \right] \nonumber \\
&=& 2\lambda T \left[ \demi + \sum_{n=1}^{\infty} \exp\left( - \frac{2\pi T \tau}{\hbar} n\right) \right] \nonumber \\
&=& \lambda T \coth \left( \frac{\pi T}{\hbar} \tau \right)
\EEA
where the integrals are re-expressed as contour integrals which are closed in the upper complex 
$\nu$-plane such that only the p\^oles with positive imaginary part will contribute. 
The last sum in the 4$^{\rm th}$ line is evaluated using  the geometric series. 
Since the result is anti-symmetric in $\tau$ (as it should be), one can continue it to the domain $\tau<0$. 

\noindent{\bf Comment:} the above derivation implicitly assumes that both times are strictly real. 
If for reasons of causality, a small imaginary part is introduced, viz.
$t-t'\mapsto t-t'+\II\vep$, one has the identity 
$\frac{1}{\tau+\II\vep} = {\rm P}\frac{1}{\tau} -\II\pi\delta(\tau)$ \cite{Gelf64}, where ${\rm P}$ denotes the 
principal part, such that (\ref{gl:B3}) would be replaced by
\BD \label{gl:B3'}
\left\langle \left\{ \eta_s(t), \eta_p(t') \right\}\right\rangle 
= 
\frac{\hbar\lambda}{\pi} \left(-\II\pi\delta(t-t') \right) 
+ \lambda T \coth \left(\frac{\pi T}{\hbar} (t-t')\right)
\tag{B.3'}
\ED
and (\ref{1.5}) would turn into 
\BEQ
\left\langle \eta_s(t)\eta_p(t')\right\rangle = \frac{\lambda T}{2}\coth\left(\frac{\pi T}{\hbar}(t-t')\right) \;\; , \;\; 
\left\langle \eta_p(t)\eta_s(t')\right\rangle 
= -\II\hbar\lambda\delta(t-t') -\frac{\lambda T}{2}\coth\left(\frac{\pi T}{\hbar}(t-t')\right) 
\EEQ

\subsection{Essential singularities and singular contributions} 

{The presentation given in the above sub-section concentrates on the situations where all variables are finite. 
However, there are certain essential singularities, which must be described in terms of singular functions(i.e. distributions).  
Reconsider the average anti-commutator 
$\langle\{\eta_s(\tau),\eta_p(0)\}\rangle = \frac{\hbar\lambda}{2\pi} J\left(\frac{\hbar\lambda}{2T},\tau\right)$. 
The integral $J(a,\tau)$ is recast into a scaling from as follows, where $a\geq 0$ is admitted throughout.  
\BEA 
\lefteqn{ J(a,\tau) \::=\: \int_{\mathbb{R}} \!\D\nu\: \coth (a\nu)\: \frac{e^{\II\nu\tau}}{\II} 
\:=\: \int_{\mathbb{R}} \!\D\nu\: \sign \nu\, \frac{\coth a|\nu|}{\nu^2} \sin \nu\tau} \nonumber \\
&=& \int_{\mathbb{R}} \!\D\nu\: \sign \nu \left[\frac{\coth a|\nu|}{\nu^2} -1 \right] \sin \nu\tau 
   + \int_{\mathbb{R}} \!\D\nu\: \sign \nu  \sin \nu\tau \nonumber \\
&=& \int_{\mathbb{R}} \!\D\nu\: \sign \nu \: \frac{e^{- a|\nu|}}{\sinh a|\nu|} \sin \nu\tau 
   + \frac{2}{\tau} \nonumber \\
&=& \frac{2}{\tau} \left( 1 + \frac{\tau}{a} \int_{0}^{\infty} \!\D x\: \frac{e^{- x}}{\sinh x} 
    \sin\left( x \frac{\tau}{a}\right)\right) \:=:\: \frac{1}{\tau} \mathscr{J}\left(\frac{\tau}{a}\right) 
\label{gl:B7-J}
\EEA
where in the third line, the singular integral was expressed as a distribution \cite[p. 174]{Gelf64}. The scaling function 
$\mathscr{J}(y)$ is well-defined for all finite values of $y$, but has an essential singularity at $y\to\infty$.}

{For the discussion of the physical properties in terms of the scaling variable $y=\tau/a$, we shall throughout fix $\tau$ and
consider the limits $a\to 0$ and $a\to\infty$, respectively. First, for $a\to\infty$, one can straightforwardly expand in $1/a$, with
the result, using \cite[(2.4.10.14)]{Prud1}  and \cite[(23.1.3)]{Abra65}
\BEA
J(a,\tau) &\simeq& \frac{2}{\tau} \left( 1 + \left(\frac{\tau}{a}\right)^2 \int_0^{\infty} \!\D x\: \frac{x\,e^{-x}}{\sinh x} 
- \frac{1}{6}\left(\frac{\tau}{a}\right)^4 \int_0^{\infty} \!\D x\: \frac{x³\,e^{-x}}{\sinh x} 
+ {\rm O}(a^{-6}) \right) 
\nonumber \\
&=& \frac{2}{\tau} \left( 1 + \frac{\pi^2}{12}\left(\frac{\tau}{a}\right)^2 
   - \frac{\pi^4}{720}\left(\frac{\tau}{a}\right)^4 + {\rm O}(a^{-6}) \right) \hfill \mbox{\rm ~;~~ $a\to\infty$}
\label{gl:B8-Jinf}
\EEA
Second, for $a\to 0$, rapid oscillations cause that the main contribution to the integral comes from around $x\approx 0$. Expanding
$e^{-x}/\sinh x\simeq x^{-1} -1 + \frac{1}{3}x + {\rm O}(x^2)$, the leading behaviour is described in terms of distributions
\BEA
J(a,\tau) &\simeq& \frac{2}{\tau} \left( 1 + \demi\frac{\tau}{a} \int_{\mathbb{R}} \!\D x\: 
\left[ \frac{1}{|x|} - 1 + \frac{|x|}{3} + \ldots \right] \sign x \sin\left( x \frac{\tau}{a}\right)\right) \nonumber \\
&=& \frac{\pi}{a} \sign \tau - \frac{2\pi}{3} a \delta'(\tau) + {\rm O}(a^2) \hfill \mbox{\rm ~;~~ $a\to 0$}
\label{gl:B9-J0}
\EEA
where again \cite[p. 174, eqs.(17,19,21)]{Gelf64} was used (the prime indicates the derivative). 
We also see that the terms of order $1/\tau$, which arose in (\ref{gl:B7-J}) as distributions, cancel. 
For the calculation of derivatives of $J(a,\tau)$, the property $\partial_{\tau}\sign\tau = 2 \delta(\tau)$ \cite[p. 22]{Gelf64}
must be used.
} 

{Similarly, we consider the integral
\BEA
\lefteqn{I(a,\tau) \::=\: \partial_{\tau}  J(a,\tau) \:=\: \int_{\mathbb{R}} \!\D\nu\: \nu\, \coth(a\nu)\, e^{\II\nu\tau} } \nonumber \\
&=& \frac{2}{\tau^2} \left( -1  
+ \left(\frac{\tau}{a}\right) \int_0^{\infty} \!\D x\: \frac{x e^{-x}}{\sinh x} \cos\left( x \frac{\tau}{a}\right)\right) 
\:=\: \frac{1}{\tau^2} \mathscr{I}\left( \frac{\tau}{a}\right) 
\nonumber \\
&\simeq& 
\left\{ \begin{array}{ll} 
      \frac{2\pi}{a} \delta\left(\tau\right) 
      - \frac{2\pi}{3} a\, \delta''\left(\tau\right) + \ldots & \mbox{\rm ~;~~ $a\to 0$} \\[0.15truecm]
      \frac{2}{\tau^2}\left( -1 + \frac{\pi^2}{12} \left(\frac{\tau}{a}\right)^2 - \frac{\pi^4}{240} \left(\frac{\tau}{a}\right)^4
      + \ldots \right) & \mbox{\rm ~;~~ $a\to\infty$} 
\end{array} \right. 
\label{gl:B10-I}
\EEA
which arises from the noise anticommutator $\left\langle\bigl\{ \zeta(\tau), \zeta(0)\bigr\}\right\rangle 
= \frac{2\hbar\lambda}{\pi m} I\left(\frac{\hbar}{2T},\tau\right)$ of the {\sc qle} (\ref{3.26}). The asymptotic forms (\ref{gl:B10-I}) 
can either be derived analogously as before
or by direct differentiation from (\ref{gl:B8-Jinf},\ref{gl:B9-J0}), 
together with the identity $\partial_{\tau}\left( |\tau|^n \sign\tau\right) = n |\tau|^{n-1}$,
for $n\in \mathbb{Z}$ and $n\ne 0$ \cite[p. 52, eq. (10)]{Gelf64}.
}

\appsection{C}{More details on noise commutators}

We analyse generic noise commutators 
\begin{equation}
\left\langle\big[\wht{\eta}_{s}(\nu),\wht{\eta}_{s}(\nu') \big] \right\rangle = \delta(\nu+\nu')\II\hbar \wht{\psi}(\nu) \;\; ,\;\;
\left\langle\big[\wht{\eta}_{p}(\nu),\wht{\eta}_{p}(\nu') \big] \right\rangle = \delta(\nu+\nu')\II\hbar \wht{\chi}(\nu)
\end{equation}
along with $\left\langle\big[\wht{\eta}_{s}(\nu),\wht{\eta}_{p}(\nu') \big] \right\rangle 
= \delta(\nu+\nu')\II\hbar \wht{\kappa}(\nu)$, instead of
(\ref{3.4}), where $\wht{\psi}$, $\wht{\chi}$ and $\wht{\kappa}$ are unknown functions. 
With the definition (\ref{3.11}), the spin-spin and momentum-momentum correlators read
\begin{align}
 \hspace{-0.3truecm}\wht{C}_{-}^{(s)}(\nu) &= \frac{\delta(\nu+\nu')\II\hbar\sqrt{\pi/2}}{(\omega^2 -\nu^2)^2 + \lambda^2 \nu^2  }
 \bigg( \frac{1}{m^2}\wht{\chi}(\nu)  
 + (\lambda^2+\nu^2)\wht{\chi}(\nu) + \frac{\II\nu+\lambda}{m}\wht{\kappa}(\nu)+ \frac{\II\nu-\lambda}{m}\wht{\kappa}(-\nu) 
 \bigg) \\
 \hspace{-0.3truecm}\wht{C}_{-}^{(p)}(\nu) &= \frac{\delta(\nu+\nu')\II\hbar\sqrt{\pi/2}}{(\omega^2 -\nu^2)^2 + \lambda^2 \nu^2  }
 \bigg(
 \nu^2\wht{\chi}(\nu)  + m^2\omega^4 \wht{\psi}(\nu) + \II m \omega^2 \nu (\wht{\kappa}(\nu) + \wht{\kappa}(-\nu) )
 \bigg)
\end{align}
Since the choice of the average noise correlators does not affect the response functions (\ref{gl:Rsp-fourier}),  
a necessary requirement of the validity of the Kubo formul{\ae}, see (\ref{gl:Kubo}), is that (\ref{3.12}) holds true, viz.  
$\wht{C}_{-}^{(p)}(\nu)\stackrel{!}{=}{m^2\omega^2} \wht{C}_{-}^{(s)}(\nu)$ (recall again (\ref{3.11})). 
This leads to the condition, for all $\nu>0$
\BEA  
& & \nu^2 \wht{\chi}(\nu) + m^2 \omega^4 \wht{\psi}(\nu) + \II m \omega^2 \nu \left( \wht{\kappa}(\nu) 
    + \wht{\kappa}(-\nu) \right) \nonumber \\
&\stackrel{!}{=}& \omega^2 \wht{\chi}(\nu) + m^2 \omega^2 \left( \lambda^2 + \nu^2\right) \wht{\psi}(\nu) +m \omega^2 
\Bigl( \II\nu \left( \wht{\kappa}(\nu) + \wht{\kappa}(-\nu) \right) 
+ \lambda\left( \wht{\kappa}(\nu) - \wht{\kappa}(-\nu) \right) \Bigr) ~~~
\EEA
The most simple solution is (which is independent of the model parameters $\omega,m$)
\BEQ
\wht{\chi}(\nu) = \wht{\psi}(\nu) = 0 \;\; , \;\; \wht{\kappa}(\nu) -\wht{\kappa}(-\nu) =0
\EEQ
Hence the form of the noise commutator correlations assumed in (\ref{3.4}), with $\wht{\kappa}(\nu)$ even, 
is the most simple formulation of a necessary condition for the validity of the Kubo formul{\ae}. 
This further implies that the commutator $\left\langle \bigl[ s(\tau), p(0)\bigr]\right\rangle$ is even in $\tau$, 
as illustrated in fig.~\ref{fig1}. 



\end{document}